\documentclass[
 reprint,
superscriptaddress,
 amsmath,amssymb,
 aps,
 prb
]{revtex4-2}

\usepackage{graphicx}
\usepackage{dcolumn}
\usepackage{bm}
\usepackage{xcolor}

\usepackage[T1]{fontenc}
\usepackage[utf8]{inputenc}
\usepackage[english]{babel}
\usepackage{stackengine}
\usepackage[]{graphicx}
\usepackage[caption=false]{subfig}
\usepackage{amsmath}
\usepackage{amssymb}
\usepackage{amsfonts}
\usepackage{times}
\usepackage{physics}
\usepackage{bbold}
\usepackage{xcolor}
\usepackage{mathtools}

\usepackage{dsfont}
\usepackage{tabularx}
\usepackage{dutchcal}

\usepackage{wasysym}

\usepackage{hyperref}
\hypersetup{colorlinks=true, citecolor=blue, urlcolor=teal, linkcolor=blue}

\usepackage{tcolorbox}
\tcbset{
    customformula/.style={
        boxrule=0.5pt,
        colback=blue!2!white, 
        colframe=blue!75!black, 
        fonttitle=\bfseries, 
        coltitle=black,
        colbacktitle=blue!5!white, 
        rounded corners=all, 
        boxsep=5pt, 
        left=0pt,
        right=5pt,
        top=5pt,
        bottom=5pt,
        before skip=10pt, 
        after skip=10pt 
    }
}

\newcommand{\oplidx}{{\text{lsp}}}
\newcommand{\Eplidx}{{\text{pl}}}
\newcommand{\zplidx}{{\text{pl}}}
\newcommand{\rplidx}{{\text{pl}}}
\newcommand{\gplidx}{{\text{pl}}}
\newcommand{\fplidx}{{\text{pl}}}
\newcommand{\pplidx}{{\text{pl}}}

\usepackage{letltxmacro}
\LetLtxMacro{\ORIGselectlanguage}{\selectlanguage}
\makeatletter
\DeclareRobustCommand{\selectlanguage}[1]{%
  \@ifundefined{alias@\string#1}
    {\ORIGselectlanguage{#1}}
    {\begingroup\edef\x{\endgroup
       \noexpand\ORIGselectlanguage{\@nameuse{alias@#1}}}\x}%
}
\newcommand{\definelanguagealias}[2]{%
  \@namedef{alias@#1}{#2}%
}

\makeatother

\definelanguagealias{en}{english}
\definelanguagealias{EN}{english}
\definelanguagealias{eng}{english}
\definelanguagealias{de}{ngerman}

\begin{document}

\preprint{APS/123-QED}

\title{Microscopic theory for a minimal oscillator model of exciton-plasmon coupling\\
in hybrids of 2d semiconductors and metal nanoparticles}

\author{Lara Greten}\email{lara.greten@tu-berlin.de}
\affiliation{Institut für Theoretische Physik, Technische Universität Berlin, Berlin, Germany}
\author{Robert Salzwedel}
\affiliation{Institut für Theoretische Physik, Technische Universität Berlin, Berlin, Germany}
\author{Diana Schutsch}
\affiliation{Institut für Theoretische Physik, Technische Universität Berlin, Berlin, Germany}
\affiliation{Present address: Department of Engineering, Universitat Pompeu Fabra, Barcelona, Spain}
\author{Andreas Knorr} \email{andreas.knorr@tu-berlin.de}
\affiliation{Institut für Theoretische Physik, Technische Universität Berlin, Berlin, Germany}
\date{\today}
\begin{abstract}
The common model to describe exciton-plasmon interaction phenomenologically is the coupled oscillator model. Originally developed for atomic systems rather than solid-state matter, this model treats both excitons and plasmons as single harmonic oscillators coupled via a constant which can be fitted to experiments.
In this work, we present a modified coupled oscillator model specifically designed for exciton-plasmon interactions in hybrids composed of two-dimensional excitons, such as in a transition metal dichalcogenide (TMDC) monolayers and metal nanoparticles while maintaining the simplicity of the commonly applied coupled oscillator models. Our approach is based on a microscopic perspective and Maxwell's equations, allowing to analytically derive an effective exciton-plasmon coupling constant.
Our findings highlight the importance of the spatial dispersion, i.e.,~the delocalized nature of TMDC excitons, necessitating the distinction between bright and momentum-dark excitons.
Both types of excitons occur at different resonance energies and exhibit a qualitatively different coupling with localized plasmons. We find a strong coupling between the plasmon and momentum-dark excitons, while a weakly coupled bright exciton manifests as an additional, third peak in the spectrum. Consequently, we propose a realistic modeling of the primary spectral features in experiments incorporating three harmonic oscillator equations instead of the conventional two. However, we also shed light on the limitations of the three coupled oscillator model in describing the line shape of extinction and scattering cross section spectra.
\end{abstract}

\maketitle

\section{Introduction}

The concept of strong coupling between semiconductor or molecular excitons and localized plasmons in metal nanostructures has become increasingly important in the field of nanophotonics, as it offers a powerful way to control and manipulate light at the nanoscale \cite{yan_2d_2020}.
Various realizations of exciton-plasmon hybrid systems exist, depending on the type of localized plasmons and excitons involved \cite{moilanen_active_2018}.

Localized surface plasmons (LSP) featured by metal nano-particles (MNPs) or lattice modes, where arrays of metal nanostructures exhibit collective plasmonic modes \cite{mahinroosta_strong_2020,salomon_strong_2013}, allow for an extreme enhancement of the electric near-field and are promising for strong coupling with excitons up to room temperature \cite{chikkaraddy_single-molecule_2016,kleemann_strong-coupling_2017,leng_strong_2018,wei_plasmonexciton_2021}.

For excitons, one distinguishes between two qualitatively different types: localized (0d) excitons, which occur in quantum dots \cite{hu_robust_2024} or molecules, e.g., J-aggregates \cite{gomez_near-perfect_2021,melnikau_strong_2013}, and two-dimensional (2d) excitons, which are delocalized across extended two-dimensional semiconductors, such as mono- or few-layers of transition metal dichalcogenides (TMDCs).
Due to the small thickness, atomically thin TMDCs exhibit high exciton binding energies thus offering stable excitonic spectral features even at room temperature. They are extremely sensitive to the environment and can be manipulated by their surrounding permittivity \cite{raja_coulomb_2017} or functionalization \cite{greten_dipolar_2024,feierabend_proposal_2017}, such as with plasmonic structures \cite{mueller_microscopic_2018,li_tailoring_2018,xie_giant_2022,low_polaritons_2017}.
Moreover, the strong excitonic dipoles and pronounced light-matter interaction \cite{kusch_strong_2021,schneider_two-dimensional_2018} of TMDCs are of particular interest and, in this context, make them suitable for studying strong coupling phenomena in hybrid MNP-excitonic systems.

In addition to electric near-field exciton-plasmon coupling, which is the focus of this work, or Purcell-enhanced emission rates \cite{kern_nanoantenna-enhanced_2015,palacios_enhanced_2017,butun_enhanced_2015}, e.g.,~for single photon generation \cite{tripathi_spontaneous_2018,von_helversen_temperature_2023,xiong_room-temperature_2021}, hybrid exciton-plasmon structures with direct electric contact \cite{stefancu_electronic_2024,li_hybrid_2016} are extensively studied regarding ultrafast charge transfer respectively hot electron injection \cite{shan_direct_2019,peng_plasmonic-hot-electron_2023}. These can be used to manipulate optical and electronic properties via doping and are particularly relevant for applications in photo-induced catalysis \cite{brongersma_plasmon-induced_2015}.

\begin{figure}[b]
    \centering
    \includegraphics[width=\linewidth]{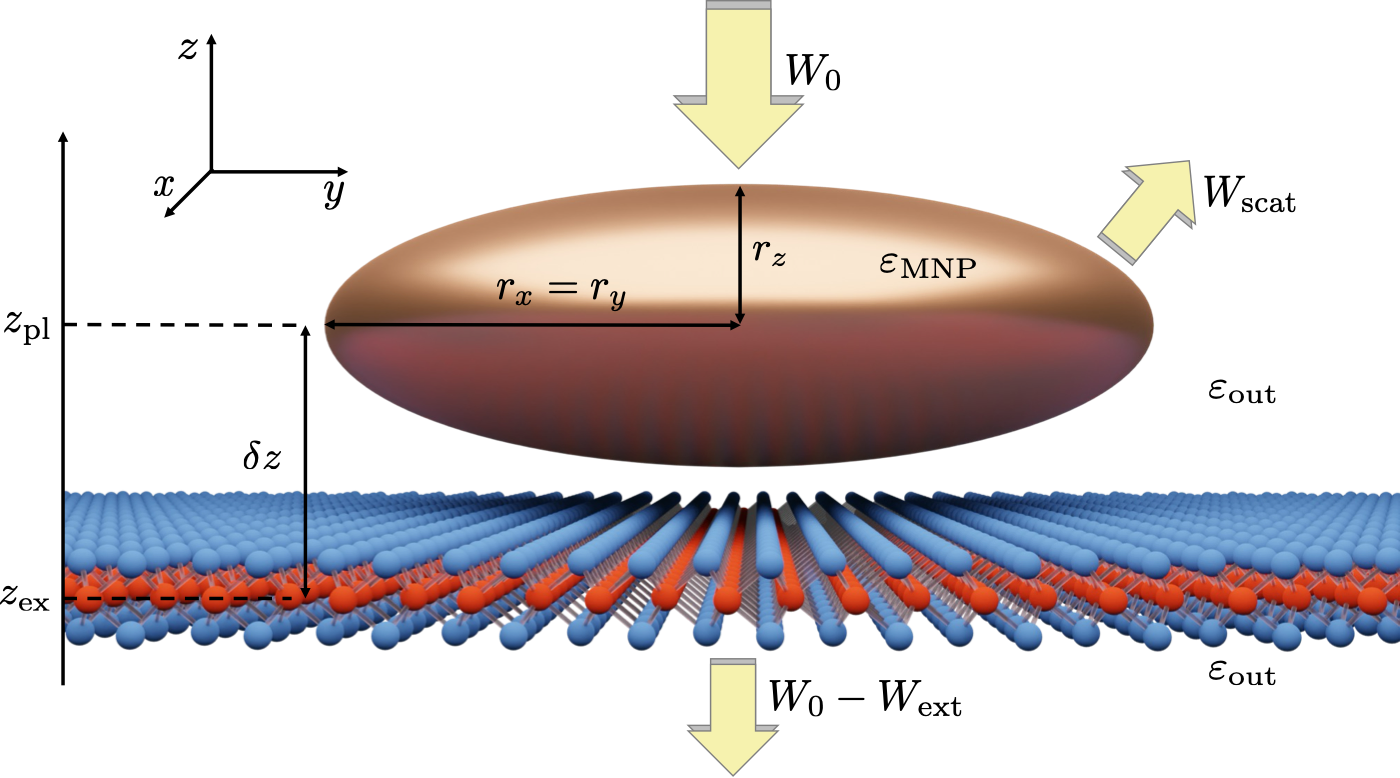}

    \caption{A gold nano-particle with half-axis $r_x$, $r_y$ and $r_z$ is located without direct contact on a TMDC monolayer with center-to-center distance $\delta z =\vert z_\zplidx -  z_\text{ex} \vert $ in a surrounding with permittivity $\varepsilon_\text{out}$. The electric field energy flow is illustrated with yellow arrows: the energy of the incident electric field $(W_0)$, the scattering $(W_\text{scat})$ and the transmitted energy $(W_0-W_\text{ext})$, that is, the incident energy minus the extinction $(W_\text{ext})$.
    }
    \label{fig:system}
\end{figure}

Figure \ref{fig:system} sketches the hybrid system discussed in this work.
It consists of a MNP on a two-dimensional TMDC monolayer substrate without direct electric contact. The MNP is modeled as a spheroid with semi-axes $r_x = r_y, r_z$ and permittivity $\varepsilon_{\text{MNP}}$ - we will use gold throughout this work - embedded in a surrounding with permittivity $\varepsilon_{\text{out}}$. The vertical separation between the MNP and the TMDC monolayer is denoted by the center-to-center distance $\delta z = \vert z_\zplidx - z_{\text{ex}}\vert $. Figure \ref{fig:system} also illustrates the incident electric field respectively its associated energy flow $W_0$ and experimentally observable quantities such as extinction $W_\text{ext}$ and scattering $W_\text{scat}$.

Since there is an increasing research interest in hybrid structures depicted in Fig.~\ref{fig:system}, models which are able to extract the coupling strength from experimental data are needed:
Common models applied to hybrid systems are the coupled oscillator model (COM) \cite{wu_quantum-dot-induced_2010,torma_strong_2014,limonov_fano_2017,joe_classical_2006}, treating the exciton and plasmon each as a harmonic oscillator, or quantum mechanically the Jaynes-Cummings-model that represents the exciton as a two-level system in a single mode electric field \cite{scully_quantum_1997,chikkaraddy_single-molecule_2016}.
In these models, the coupling can be classified into different regimes. In the weak coupling regime, the interaction primarily enhances decay rates, such as the exciton’s radiative decay. In the strong coupling regime, when the coupling strength exceeds the decay rates, energy is coherently exchanged between the two modes over several oscillations, leading to the formation of new hybridized states, observable as an effective Rabi splitting in the spectrum \cite{torma_strong_2014}. More extreme regimes, such as ultra-strong \cite{wu_ultrastrong_2024} and deep-strong \cite{mueller_deep_2020} coupling, involve even stronger interactions, where the coupling strength becomes a significant fraction of or comparable to the mode frequency. Since this leads to the breakdown of several approximations \cite{hughes_reconciling_2024,forn-diaz_ultrastrong_2019,frisk_kockum_ultrastrong_2019}, we will not delve into these extreme regimes here.

When these models are applied to measurements of strong coupling for plasmonic structures interacting with localized (0d) excitons such as in Refs.~\cite{gomez_near-perfect_2021,todisco_excitonplasmon_2015,wang_interplay_2014,wu_plexcitonic_2021,chikkaraddy_single-molecule_2016,mahinroosta_strong_2020,todisco_ultrastrong_2018} or 2d TMDC excitons \cite{abid_temperature-dependent_2017,bisht_collective_2019,lee_electrical_2017,zhang_observation_2023,abid_resonant_2016,liu_strong_2016,wang_coherent_2016,zhu_electroluminescence_2023,petric_tuning_2022,zheng_manipulating_2017,cuadra_observation_2018,geisler_single-crystalline_2019,kleemann_strong-coupling_2017,qin_revealing_2020,wen_room-temperature_2017}, the coupling constant is treated as a fit parameter to categorize hybrid systems into a coupling regime.
While this approach is useful for categorizing experimental data, it provides limited physical insight, particularly for the more complex behavior of 2d semiconductor excitons, which involve momentum-dark and bright states, typically not distinguished in phenomenological COMs. Computational simulations via Maxwell solvers are often applied \cite{bisht_collective_2019,zhu_electroluminescence_2023,lee_fano_2015,zheng_manipulating_2017,pincelli_observation_2023,cuadra_observation_2018,zhang_observation_2023,geisler_single-crystalline_2019,kleemann_strong-coupling_2017,petric_tuning_2022} for more quantitative predictions, but their quality depends on the applied excitonic model.    

The goal of this paper is to propose an analytical, physically grounded model, which includes specifically important features of 2d excitons beyond a simple oscillator model. We aim to provide a model that can be easily applied to experimental data, while also offering a clear physical understanding of the underlying coupling mechanisms and incorporating key extensions to the COM to account for the spatial delocalization of the 2d excitons.
    
The paper is structured as follows: Section \ref{sec: basics of excitons and plasmons} provides a theoretical background on the optical properties of MNPs and two-dimensional excitons, e.g.,~in TMDC monolayers. It introduces the TMDC circular dichroism and, due to the delocalized nature of the 2d excitons, distinguishes between bright and momentum-dark excitons on the exciton dispersion.
Starting from the microscopic exciton and plasmon dynamics, we derive in Sec.~\ref{sec: Exciton-Plasmon Coupling} a coupled harmonic oscillator model for the exciton-plasmon interactions
and provide
formulas for typical, experimentally accessible observables.
In this context, the section delves into the strongly different coupling character of bright and momentum-dark excitons to plasmons.
We provide explicit microscopic parameters, which determine all coupling and dephasing constants in a COM combining three oscillators: plasmons, momentum-dark and bright excitons.
In the results Sec.~\ref{sec: results}, we evaluate the observables under various conditions, demonstrating the sensitivity of the coupling strength on the exciton-plasmon distance, as well as to temperature and MNP packing density to highlight how the developed model differs from to the conventional two coupled oscillator model.

\section{Optical Properties of Excitons and Plasmons \label{sec: basics of excitons and plasmons}}
In this section, we introduce excitons in TMDC monolayers via the excitonic Bloch equation \cite{kira_semiconductor_2012} and discuss the distinction between bright and momentum-dark excitons. Next, we present the modeling of the optical near- and far-field response of plasmons in MNPs using Mie-Gans theory \cite{mie_beitrage_1908,gans_uber_1912,bohren_absorption_1983}. In both cases, we derive harmonic oscillator equations from the microscopic theory to approximately model their dynamics as driven by the incident electric field and occurring near-fields.
MNP plasmon and the TMDC excitons are, if no electronic overlap (responsible for tunneling) exists, coupled
via the total electric near-field \cite{greten_strong_2024} as the sum of the fields generated by exciton $\mathbf{E}^\text{ex}$, plasmon $\mathbf{E}^\Eplidx$ and the incident electric field $\mathbf{E}^0$:
\begin{align}
   \mathbf{E}(\mathbf{r},\omega) = \mathbf{E}^\text{ex}(\mathbf{r},\omega)+\mathbf{E}^\Eplidx(\mathbf{r},\omega)+\mathbf{E}^0(\mathbf{r},\omega) \label{eq: total electric field 0}
\end{align}
All contributions in Eq.~\eqref{eq: total electric field 0} are determined as a solution of Maxwell's equations.

\subsection{Excitons \label{sec: basics excitons}}
\begin{figure}[t]
    \includegraphics[width=\linewidth]{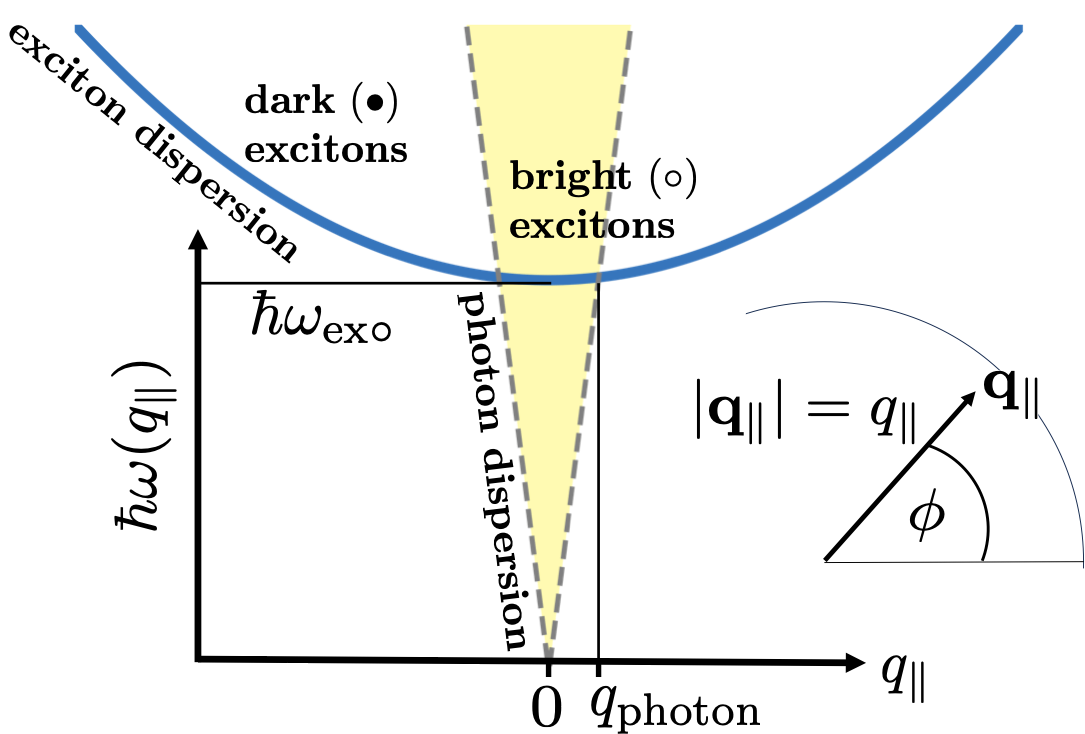}
    \caption{Sketch of the free exciton dispersion $\hbar\omega_\text{ex}( q_\parallel)$ (blue). The light-cone, within the photon dispersion $\hbar\omega_\text{photon}= \frac{c}{\varepsilon_\text{out}} q$, is shaded in yellow.
    Excitons in the light-cone ($q_\parallel < q_\text{photon}$) are bright ($\circ$).
    Excitons out-side the light-cone ($q_\parallel > q_\text{photon}$) are momentum-dark ($\bullet$).
    \textbf{inset:} 2d momentum $\mathbf{q}_\parallel$ in polar representation with absolute value $\vert \mathbf{q}_\parallel\vert =q_\parallel$ and angle $\phi$.}    
     \label{fig: free exciton dispersion sketch}
\end{figure}

The \textit{Exciton Bloch Equation} for the dominant 1s excitonic transition $p^{\xi}_\mathbf{q_\parallel}$ with in-plane Fourier component $\mathbf{q_\parallel}$ (also referred to as in-plane momentum) reads \cite{greten_dipolar_2024}
\begin{align}
\left(\hbar\omega
-\hbar \omega_\text{ex}(\mathbf{q_\parallel})+\frac{i}{2}\hbar\gamma^\text{ex}(T) \right) p^{\xi}_\mathbf{q_\parallel}(\omega)
= 
- \varphi_{0}^* {\mathbf{d}^{\xi}}^* \cdot \mathbf{E}_{\mathbf{q}_\parallel}(\omega)\label{eq:excitonicBlochEquation}
\end{align}
Equation \eqref{eq:excitonicBlochEquation} restricts to the lowest, 1s excitonic state, since it is energetically separated from higher excitonic transitions~\cite{wang_colloquium_2018}.
In Eq.~\eqref{eq:excitonicBlochEquation}, $\xi= K^+/K^-$ denotes the valley with direct band gaps, located at the corners of the first Brillouin zone of the electronic band structure.
In the vicinity of the $K^+/K^-$ valleys, the momentum-dependent exciton resonance frequency is described in parabolic approximation \cite{knorr_theory_1996,selig_ultrafast_2019}
\begin{align}
   \hbar \omega_\text{ex}(\mathbf{q_\parallel}) = \hbar \omega_{\text{ex}\circ}+ \frac{\hbar^2\mathbf{q}_\parallel^2}{2M}, \label{eq: parabolic exciton dispersion}
\end{align}
with the effective electron mass $M$ parameterized from DFT calculations \cite{kormanyos_k_2015} and the resonance frequency $\omega_{\text{ex}\circ}$ as observed in optical far-field spectroscopy. $\mathbf{q_\parallel}\approx 0$ corresponds to bright excitons, all other excitons are optically dark \cite{selig_dark_2017,katzer_impact_2023}. Exciton and photon dispersions are schematically displayed in Fig.~\ref{fig: free exciton dispersion sketch}.

The temperature-dependent damping $\gamma^\text{ex}(T)$ in Eq.~\eqref{eq:excitonicBlochEquation} accounts for non-radiative decay due to exciton-phonon interactions and is determined microscopically in Ref.~\onlinecite{selig_excitonic_2016}
\begin{align}
    \hbar\gamma^\text{ex}(T) = c_1 T+\frac{c_2}{e^{\frac{\Omega}{k_BT}}-1}. \label{eq: exciton damping definition (1. order)}
\end{align}
All parameters, including $c_1$, $c_2$, and the average phonon energy $\Omega$, are listed in table \ref{table: parameters}.
The parameter $\varphi_{0}$ is the value of the 1s excitonic wave function at the origin in real space \cite{berghauser_analytical_2014,selig_excitonic_2016}.
The valley-dependent dipole element $\mathbf{d}^\xi$ with $\mathbf{d}^{K^+}\cdot \mathbf{d}^{K^-} =0$ and absolute value $\vert \mathbf{d}^\xi\vert = d$ is derived from DFT calculations in Ref.~\onlinecite{xiao_coupled_2012}.

\subsubsection{Exciton dipole density as harmonic oscillator}
The macroscopic 2d exciton dipole density resulting from the excitonic transition is \cite{haug_quantum_2004}
\begin{align}
{\mathbf{P}^{\text{ex}}_{\mathbf{q_\parallel}}}
&=\sum_\xi \varphi_{0} \, \mathbf{d}^\xi\, p^{\xi}_{\mathbf{q_\parallel}}  + \,{ h.c.} \label{eq: exciton dipole density, quantum definition}
\end{align}
To achieve an oscillator representation often used in phenomenological models of hybrid MNP-TMDC structures, we apply the Heisenberg equation formalism twice to Eq.~\eqref{eq: exciton dipole density, quantum definition}, similar to Ref.~\onlinecite{schubert_nonlinear_1986} and assume that the damping is small compared to the resonance frequency $({\gamma^\text{ex}})^2\ll \omega_{\text{ex}\circ}^2$.
For a real-valued electric field $\mathbf{E}_{\mathbf{q}_\parallel}(t)$ in the time domain, we find the following second-order harmonic oscillator for the in-plane components of the dipole density $\mathbf{P}^{\text{ex}}_\mathbf{q_\parallel}$:
\begin{align}
\left(\omega^2 +i\gamma^\text{ex}(T)\omega - \omega_\text{ex}^2(q_\parallel)\right)\mathbf{P}^{\text{ex}}_\mathbf{q_\parallel}
=-\frac{2\omega_\text{ex}({q_\parallel})}{\hbar}\, \vert\varphi_{0} d\vert^2\mathbf{E}_{\mathbf{q_\parallel}}(\omega)\label{eq:macroscopic exciton second order}.
\end{align}
The $z$ component of $\mathbf{P}^{\text{ex}}$ is negligible \cite{slobodeniuk_spinflip_2016,wang_-plane_2017} compared to the in-plane dipole density due to the small thickness of the TMDC.

\subsubsection{Momentum-Bright Excitons}
The dispersion $\omega_\text{ex}(\mathbf{q}_\parallel)$ describes a variety of exciton states with quantum number $\mathbf{q}_\parallel$.
Whether an exciton state with $\omega_\text{ex}(\mathbf{q}_\parallel)$ is referred to as bright $(\circ)$ or dark $(\bullet)$ is characterized by its accessibility under far-field illumination. The corresponding selection rules apply to energy, spin \cite{zhou_probing_2017}, or, as relevant in this case, to momentum $\mathbf{q}_\parallel$.
The photon momentum reads ${q}_\text{photon} = \sqrt{\varepsilon_\text{out}} \frac{ \omega}{c}$ and, due to in-plane momentum ($\mathbf{q_\parallel}$) conservation for the excitation process, bright excitons can only possess comparably small in-plane momenta $q_\parallel$:
\begin{equation}
     q^2_\parallel < \varepsilon_\text{out} \frac{ \omega^2}{c^2} \label{eq: def in light cone}.
\end{equation}
Eq.~\eqref{eq: def in light cone} refers to momenta within the light cone.
Due to their small momenta, bright excitons are spatially delocalized over large areas in real space (comparable with the wavelength of the incoming light) fulfilling Abbe's diffraction limit.
To derive a phenomenological model to fit experiments, we qualitatively consider the $q_\parallel=0$ exciton as a representative mode for all excitons within the light cone since $\omega_\text{ex}({q_\parallel}< \varepsilon_\text{out} \frac{ \omega^2}{c^2})\approx \omega_{\text{ex}\circ}$ is valid for frequencies $\omega$ in the visible range:
\begin{align}
\left( \omega^2 +i\gamma^\text{ex}_\circ(T)\omega - \omega_{\text{ex}\circ}^2\right) \mathbf{P}^{\text{ex}\circ}_\mathbf{q_\parallel=0}
=-\frac{2\omega_{\text{ex}\circ}}{\hbar}\, \vert\varphi_{0} d\vert^2\mathbf{E}_{\mathbf{q_\parallel=0}}, \label{eq: 1. bright exciton HO}
\end{align}
depending on the electric field, Eq.~\eqref{eq: total electric field 0}.
However, we already accounted for the self-interaction $\mathbf{E}^\text{ex}$, which gives rise to radiative damping \cite{selig_excitonic_2016}, second term in Eq.~\eqref{eq: damping bright exciton}:
\begin{align}
\gamma^\text{ex}_{\circ}(T) = \gamma^\text{ex}(T)
+\vert\varphi_{0} d\vert^2 \frac{\omega_\text{ex}}{\hbar c \sqrt{\varepsilon_\text{out}}\varepsilon_0}. \label{eq: damping bright exciton}
\end{align}
In Eq.~\eqref{eq: 1. bright exciton HO} we add the symbol $\circ$ to indicate the brightness of the excitons.

\subsubsection{Momentum-Dark Excitons}
Momentum-dark excitons ($\bullet$), on the other hand, have larger in-plane momenta,
\begin{equation}
     q^2_\parallel > \varepsilon_\text{out} \frac{ \omega^2}{c^2} \label{eq: def out light cone},
\end{equation}
positioning them outside the light cone and thus making them optically inaccessible under far-field excitation.

Momentum-dark exciton distributions may have a larger extent in momentum space, enabling them to be spatially localized in real space. The correspondingly generated electric field gradients lead to a momentum-dependent modification of their dispersion:
\begin{align}
&\left( \omega^2 +i\gamma^\text{ex}\omega - \omega_\text{ex}^2(q_\parallel)\right) \mathbf{P}^{\text{ex}\bullet}_\mathbf{q_\parallel}(\omega)\label{eq: 1. dark exciton HO}\\
&=\frac{\omega_\text{ex}(q_\parallel)}{\hbar}\, \vert\varphi_{0} d\vert^2 \frac{1}{\varepsilon_0\varepsilon_\text{out}}
q_\parallel \mathbcal{U}_\phi\cdot
{\textbf{P}^{\text{ex}\bullet}_{\mathbf{q_\parallel}}}-\frac{2\omega_\text{ex}(\mathbf{q_\parallel})}{\hbar}\, \vert\varphi_{0} d\vert^2\mathbf{E}_{\mathbf{q_\parallel}}\nonumber
\end{align}
where the symbol $\bullet$ indicates the momentum-dark character of the excitons and with the degenerate, idempotent matrix
\begin{equation}
   \mathbcal{U}_\phi = \frac{\mathbf{q}_\parallel\otimes\mathbf{q}_\parallel}{q_\parallel^2} =  \left( \begin{array}{cc} 
\cos^2 (\phi) &   \cos (\phi)\sin (\phi)\\ 
\cos (\phi)\sin (\phi) & \sin^2 (\phi)  \\ 
\end{array}\right)
\label{eq: matrix U}
\end{equation}
depending on the momentum $\mathbf{q}_\parallel$ in polar representation $({q}_\parallel,\phi)$, cp.~Fig.~\ref{fig: exciton dispersion}.
Similar to the bright excitons, the dark-exciton dynamics are driven by the electric field, Eq.~\eqref{eq: total electric field 0}, without $\mathbf{E}^\text{ex}$ as we have singled out the electric field-mediated self-interaction, given by Eq.~\eqref{eq: matrix U}.
Despite being optically inaccessible from the far-field, dark excitons may play a crucial role in near-field interactions as we will demonstrate in this manuscript for the context of strong coupling with plasmonic nano-structures.

\subsection{Localized Surface Plasmons \label{sec: basics plasmon}}
The response of a metal nano-particle with extensions small compared to the incident wavelength, with spheroidal shape and permittivity $\varepsilon_\text{MNP}$ can be described as a point dipole $\mathbf{p}^\pplidx $ by Mie-Gans theory \cite{mie_beitrage_1908,gans_uber_1912,bohren_absorption_1983}. In quasi-static approximation, it is given by a polarizability $\boldsymbol{\alpha}$ times the total electric field, cp.~Eq.~\eqref{eq: total electric field 0}, again by excluding the electric field generated by the MNP, i.e.,~the self interaction.
\begin{equation}
\mathbf{p}^\pplidx (\omega) = \boldsymbol{\alpha}(\omega)\cdot \mathbf{E}(\mathbf{r}_\rplidx ,\omega) \label{eq: plasmon 0}
\end{equation}
The polarizability of an oblate metal nano-spheroid is, in a Cartesian basis, diagonal and we choose $z$ to be along the short semi-axis, cp.~Fig.~\ref{fig:system}, such that its entries are $\alpha_{x} = \alpha_y \neq \alpha_z$ with
\begin{equation}
{p}^\pplidx _j(\omega) = {\alpha}_j(\omega) {E}_j(\mathbf{r}_\rplidx ,\omega) ,\label{eq: single nano-particle dipole}
\end{equation}
and 
\begin{align}
    \alpha_j = V_\text{MNP}\,  \varepsilon_0\varepsilon_\text{out}
    \frac{\varepsilon_\text{MNP}-\varepsilon_\text{out}}{L_j \varepsilon_\text{MNP}+(1-L_j)\varepsilon_\text{out}}
    \label{eq: plasmon alpha}.
\end{align}
$V_\text{MNP}$ is the volume of the spheroidal MNP
\begin{align}
    V_\text{MNP} = \frac{4}{3} \pi \, r_x^2 r_z
\end{align}
and
the shape parameters $L_j$ are given by \cite{bohren_absorption_1983}
\begin{align}
\quad L_z &= \frac{1}{e_\text{MNP}^2}\qty(1-\frac{\sqrt{1-e_\text{MNP}^2}}{e_\text{MNP}}\arcsin(e_\text{MNP}))\label{eq:lz}\\
\text{and}
\quad L_x &=L_y = \frac{1-L_z}{2},\label{eq:lxly}
\end{align}
with the spheroids eccentricity 
\begin{align}
e_\text{MNP}&=1-\frac{r_z^2}{r_{x}^2}.
\end{align}
In addition to the geometry, the polarizability $\boldsymbol{\alpha}$, Eq.~\eqref{eq: plasmon alpha}, depends on the permittivity $\varepsilon_\text{out}$ of the surrounding, cp.~Fig.~\ref{fig:system}, and the permittivity of the MNP itself
\begin{align}
\varepsilon_\text{MNP}(\omega)  =&\,\varepsilon_b(\omega) + \chi_\text{d}(\omega) \label{eq: MNP permittivity}.
\end{align}
The Drude susceptibility, $\chi_\text{d}$ describes intraband transitions for the quasi-free electrons in the partially occupied metal conduction band. \cite{landau_vibrations_1945,drude_zur_1900}
\begin{align}
\chi_d(\omega)= \frac{-\omega_\text{pl,bulk}^2}{\omega^2+i\omega\gamma^\text{pl}(\omega,T)} \label{eq: Drude susceptibility}
\end{align}
with the bulk plasma frequency $\omega_\text{pl,bulk}$ and the plasmonic damping $\gamma^\text{pl}(\omega,T)$.
All other contributions to the MNP susceptibility $\varepsilon_\text{MNP}$ are summarized in the MNP background permittivity
\begin{align}
\varepsilon_b(\omega) =&\,\varepsilon_\infty + \chi_\text{inter}(\omega) \label{eq: MNP background permittivity}.
\end{align}
These include the high-frequency limit dielectric constant $\varepsilon_\infty$ stemming from inner bands and, depending on the chosen metal, frequency dependent inter-band transitions between valence and conduction band in $\chi_\text{inter}(\omega)$. In gold, these transitions become significant above $2.4\,\text{eV}/\hbar$ \cite{wang_cu_2005}. Although the precise location of visible interband transitions within the gold band structure remains a subject of debate, several models \cite{vial_improved_2005,etchegoin_analytic_2006,myroshnychenko_modelling_2008,kreibig_optical_2013} have been developed that align well with experimental observations provided in refs.~\onlinecite{johnson_optical_1972,palik_handbook_1998}. For this work, we adopt the interband susceptibility described in Ref.~\onlinecite{etchegoin_analytic_2006}, given by a sum of Lorentzian functions 
\begin{align}
    \label{eq:permittivity}
    \chi_\text{inter}(\omega) = 
    &\sum_{j = 1,2}
        A_j\omega_j
        \biggl[
            \frac{e^{i\phi_j}}{\omega_j -\omega -i\Gamma_j}
            +
            \frac{e^{-i\phi_j}}{\omega_i +\omega +i\Gamma_j}
        \biggr]
\end{align}
In the Drude contribution, Eq.~\eqref{eq: Drude susceptibility}, we use a damping rate $\gamma^\text{pl}(\omega,T)$ that depends on both temperature and frequency \cite{liu_reduced_2009,mckay_temperature_1976,parkins_intraband_1981,bouillard_low-temperature_2012}. It accounts for electron-electron and electron-phonon scattering:
\begin{align}
    \gamma^\text{pl} (\omega,T) = \gamma^\text{pl}_{\text{el-el}}(\omega,T) + \gamma^\text{pl}_{\text{el-ph}}(T),\label{eq: drude damping}
\end{align}
with~\cite{liu_reduced_2009,mckay_temperature_1976,parkins_intraband_1981,bouillard_low-temperature_2012}
\begin{align}
    \gamma^\text{pl}_{\text{el-el}}(\omega,T) &=  \frac{b}{\hbar} \left[ (k_BT)^2 + (\hbar \omega/2\pi)^2\right]\label{eq: drude damping - elel},\\
    \gamma^\text{pl}_{\text{el-ph}}(T) &=  \frac{\gamma_0}{\hbar}\left[ \frac{2}{5}+ 4\left(\frac{T}{\Theta}\right)^5\int_0^{\Theta/T} \frac{z^4}{e^z-1}dz\right].\label{eq: drude damping - elph}
\end{align}

 Similar to the description of the exciton oscillator, Eqs.~(\ref{eq: 1. bright exciton HO},\ref{eq: 1. dark exciton HO}), we combine Eqs.~\eqref{eq: single nano-particle dipole}-\eqref{eq: Drude susceptibility} to obtain an equation for each entry of the plasmon dipole. We assume $\omega E_j(\mathbf{r}_\rplidx )\approx 0$ to be consistent with Eq.~\eqref{eq: single nano-particle dipole} that is defined in the quasi-static limit.
\begin{align}
    & \left( \omega^2 + i\gamma^\text{pl}(\omega,T) \omega - \frac{\omega_\text{pl,bulk}^2 L_j}{L_j \varepsilon_b(\omega)  + (1-L_j) \varepsilon_\text{out}} \right) p^\pplidx_j(\omega)\nonumber\\
    &=
    -\frac{\omega_\text{pl,bulk}^2 \varepsilon_0\varepsilon_\text{out}V_\text{MNP}}{L_j \varepsilon_b(\omega)  + (1-L_j) \varepsilon_\text{out}} E_j(\mathbf{r}_\rplidx ,\omega) \label{eq: HO plasmon}
\end{align}
To obtain a pure classical harmonic oscillator model from Eq.~\eqref{eq: HO plasmon}, we approximate $\varepsilon_b(\omega)\approx \Re\varepsilon_b(\omega_\oplidx)\equiv \varepsilon_b$, $ \gamma^\text{pl}(\omega,T)
\approx
\gamma^\text{pl}(\omega_\oplidx,T)
\equiv
\gamma^\text{pl}(T) $ by dropping the respective frequency dependencies in the relevant frequency range with $E_j(\omega;\mathbf{r}_\rplidx )\neq 0$ which we therefore restrict to the frequency regime where $\hbar\omega < 2.4$\,eV \cite{wang_cu_2005}.
This approximation is necessary to treat the plasmon dipole as a single harmonic oscillator, and it resembles the main weakness of this simplified model for the MNP dynamics. Thus, the inner metal shells and interband transition contributions to the permittivity are treated parametrically in the macroscopic Maxwell equations, therefore altering the electric field solution. Implications of this approximation are discussed in Appendix \ref{app sec: permittivity dependency}.
In the case of the spheroid, the geometry breaks the symmetry between the in-plane ($x/y$) and $z$-direction. However, the coupling to the TMDC excitons is dominated by the $x/y$-components of the MNP dipole that is chosen to be resonant or near-resonant to the TMDC excitons and orientated parallel to the TMDC plane \cite{mueller_microscopic_2018}, cp.~Fig.~\ref{fig:system}. We provide a detailed justification for dropping the $z$-direction in Sec.~\ref{sec: Exciton-Plasmon Coupling} and Appendix \ref{sec: Appendix to 2 dim}.
All in all, Eq.~\eqref{eq: HO plasmon} simplifies to a single harmonic oscillator equation for the relevant (in-plane) plasmon dipole contribution $\mathbf{p}^\pplidx $:
\begin{align}
    \left( \omega^2 +i \gamma^\text{pl}(T) \omega  -\omega_{\oplidx}^2 \right) \mathbf{p}^\pplidx (\omega)
    &=-V_\text{MNP}\frac{\varepsilon_0\varepsilon_\text{out} }{L}
    \omega_{\oplidx}^2 \mathbf{E}(\omega;\mathbf{r}_\rplidx ) \label{eq: plasmon equation}
\end{align}
with the localized surface plasmon (LSP) frequency of the MNP,
\begin{align}
    \omega_{\oplidx} = \sqrt{ \frac{\omega_\text{pl,bulk}^2 L}{L \varepsilon_b  + (1-L) \varepsilon_\text{out}}},
\end{align}
which differs from the bulk plasma frequency $\omega_\text{pl,bulk}$ due to the influence of the MNP geometry via the shape factor $L$, Eq.~\eqref{eq:lxly}, and the background $\varepsilon_b$ as well as the surrounding permittivity $\varepsilon_\text{out}$.

\section{Exciton-Plasmon Coupling \label{sec: Exciton-Plasmon Coupling}}
Now we turn to the description of the exciton-plasmon coupling. Exciton and plasmon interact via the total electric field, cp.~Eq.~\eqref{eq: total electric field 0}, such that the previously derived individual harmonic oscillator Eqs.~(\ref{eq: 1. bright exciton HO},\ref{eq: 1. dark exciton HO},\ref{eq: plasmon equation}) couple. When inserting the occurring electric fields, we have to consider, that the self-interaction generated by the MNP and TMDC respectively has already been self-consistently taken into account in their individual harmonic oscillator dynamics.
Considering the cross-field interaction via $\mathbf{E}^\text{ex}$ and $\mathbf{E}^\Eplidx$ and the incident field $\mathbf{E}^0$, we obtain a set of coupled harmonic oscillator equations -
the bright exciton oscillator equation
\begin{align}
 &\left( \omega^2 +i\gamma^\text{ex}_\circ\omega - \omega_{\text{ex}\circ}^2\right) \mathbf{P}^{\text{ex}\circ}_{\mathbf{q_\parallel}=\mathbf{0}}(\omega)\label{coupled eq 1: bright exciton}\\
&= -\frac{2\omega_{\text{ex}\circ}}{\hbar}\, \vert\varphi_{0} d\vert^2\left(
 \mathbf{E}^\Eplidx_{\mathbf{q_\parallel}=\mathbf{0}}(z_\text{ex};\omega)
+
\mathbf{E}^0_{\mathbf{q_\parallel}=\mathbf{0}}(z_\text{ex};\omega)
\right)
,\nonumber
\end{align}
one oscillator equation for each momentum-dark exciton, with momenta ${q_\parallel}>\sqrt{\varepsilon_\text{out}} \frac{ \omega}{c}$,
\begin{align}
&\left( \left( \omega^2 +i\gamma^\text{ex}\omega - \omega_\text{ex}^2(q_\parallel)\right)\mathds{1} 
-\frac{\omega_\text{ex}(q_\parallel)}{\hbar}
\frac{\vert\varphi_{0} d\vert^2}{\varepsilon_0\varepsilon_\text{out}}
q_\parallel \mathbcal{U}_\phi\right)\cdot
{\textbf{P}^{\text{ex}\bullet}_{\mathbf{q_\parallel}}}(\omega)\nonumber\\
&=
-\frac{2\omega_\text{ex}(q_\parallel)}{\hbar}\, \vert\varphi_{0} d\vert^2
\left(
 \mathbf{E}^\Eplidx_{\mathbf{q_\parallel}}(z_\text{ex};\omega)
+
\mathbf{E}^0_{\mathbf{q_\parallel}}(z_\text{ex};\omega)
\right)
\label{coupled eq 1: dark exciton},
\end{align}
and the plasmon oscillator equation
\begin{align}
    &\left( \omega^2 + i\gamma^\text{pl}  \omega - \omega_{\oplidx}^2 \right) \mathbf{p}^\pplidx (\omega)\nonumber\\
    &=-V_\text{MNP}\frac{\varepsilon_0\varepsilon_\text{out} }{L}
    \omega_{\oplidx}^2
    \left(
    \mathbf{E}^\text{ex}(\mathbf{r}_\rplidx ;\omega) + \mathbf{E}^0(\mathbf{r}_\rplidx ;\omega) 
    \right)
    \label{coupled eq 1: plasmon}.
\end{align}

The electric field emitted by the TMDC excitons at an arbitrary position $z$ is expressed via the dyadic Green's function $\mathbcal{G}$ and transformed to real space, similar to Ref.~\onlinecite{greten_dipolar_2024}. 
Without loss of generality, we set the in-plane coordinate of the MNP to $\mathbf{r_{\parallel \text{MNP}}}=\mathbf{0}$. Expressing the fields by their sources, we have:
\begin{align}
\mathbf{E}^{\text{ex}\circ/\bullet}(\mathbf{r}_\rplidx ;\omega)
&= \frac{1}{(2\pi)^2} \int d^2 q_\parallel
\mathbcal{G}_{\mathbf{q_\parallel}}(z_\zplidx,z_\text{ex};\omega)\cdot {\mathbf{P}^{\text{ex}\circ/\bullet}_{\mathbf{q_\parallel}}}(\omega). \label{eq: electric field of TMDC at MNP}
\end{align}
Analogous, the electric field emitted by the MNP
in momentum space and observed in the TMDC layer at $z_\text{ex}$ is given by
\begin{align}
\mathbf{E}^{\Eplidx}_\mathbf{q_\parallel}(z_\text{ex};\omega)= \mathbcal{G}_{\mathbf{q_\parallel}}(z_\text{ex},z_\zplidx;\omega)\cdot \mathbf{p}^\pplidx (\omega). \label{eq: electric field of MNP at TMD}
\end{align}
For an incident field $\mathbf{E}^0_\mathbf{q_\parallel}\delta_{\mathbf{q_\parallel},\mathbf{0}}$ with only in-plane electric field components, such as a plane wave propagating perpendicular to the TMDC, neither the TMDC nor the MNP polarization gain a contribution in $z$-direction. Therefore, we may drop all $z$-components in the following, which reduces the effective dimension of Eqs.~(\ref{coupled eq 1: bright exciton}-\ref{coupled eq 1: plasmon}) each to two. We provide a formal proof that this effective $2$d theory in fact resembles the coupling in the $3$d hybrid structure in Appendix \ref{sec: Appendix to 2 dim}. Thus, without approximation, the dyadic Green's function reduces to $2\times 2$ with
\begin{align}
&\mathbcal{G}_{{\mathbf{q_\parallel}}}(z,z',\omega) =
\begin{pmatrix}
-\frac{\omega^2}{{\varepsilon}_0c^2}\mathbb{1} + \frac{{\mathbf{q_\parallel}}\otimes\,{\mathbf{q_\parallel}}}{{\varepsilon}_0\varepsilon_\text{out}}\\
\end{pmatrix}
G_{{q_\parallel}} (z,z',\omega). \label{eq:dyadic Green's function}
\end{align}
For simplicity, we apply the scalar Green's function for a homogeneous dielectric environment
\begin{align}
G_{q_\parallel}(z,z',\omega)&=\frac{-i}{2k_{q_\parallel}}e^{ik_{q_\parallel}|z-z'|}\label{eq: Green's function, omega dependent}
\end{align}
with the wavevector $k_{q_\parallel}=\sqrt{\frac{{\varepsilon_\text{out}}}{c^2}\omega^2-q_\parallel^2}$.

Combining Eqs.~(\ref{coupled eq 1: plasmon}, \ref{eq: electric field of MNP at TMD} and \ref{eq:dyadic Green's function}) shows that the MNP provides high $\mathbf{q}_\parallel$ (evanescent) electric fields, that allow to access originally momentum-dark excitons in the TMDC, cp.~Eq.~\eqref{coupled eq 1: dark exciton}.
Furthermore, from Eqs.~(\ref{eq: plasmon 0}-\ref{eq: plasmon alpha}) it follows that, for a plane-wave excitation propagating along $z$, the orientation of the MNP dipole $\mathbf{p}^\pplidx $ directly follows the orientation of the electric field $\mathbf{E}(\mathbf{r}_\rplidx )$. Consequently, a circular $(\sigma)$ polarized incident electric field $\mathbf{E}^0$ yields $\mathbf{p}^\pplidx \parallel \mathbf{e}_{\sigma}$.
However, a circularly polarized oriented point dipole does not generate a uniformly circular polarized electric near-field, cp.~Eqs.~(\ref{eq: electric field of MNP at TMD},\ref{eq:dyadic Green's function}). In contrast, its electric near-field possesses contributions in both circular polarization directions, depending on the relative position to the MNP.
Consequently, even for circularly polarized excitation, the MNP acts as an excitation and couples to excitons in both valleys $K^+$ and $K^-$. \cite{salzwedel_spatial_2023} 
In the next three sections (\ref{sec: quasistatic coupling},\ref{sec: coupling of the plasmons} and \ref{sec: bright excitons}) we simplify the action of the interacting fields in the coupled equations (\ref{coupled eq 1: bright exciton},\ref{coupled eq 1: dark exciton},\ref{coupled eq 1: plasmon}) to obtain a simplified system of three coupled oscillators, which can be used to fit experiments.

\subsection{Coupled oscillator equation for momentum-dark excitons \label{sec: quasistatic coupling}}
\begin{figure}[t]
    \centering
    \includegraphics[width=\linewidth]{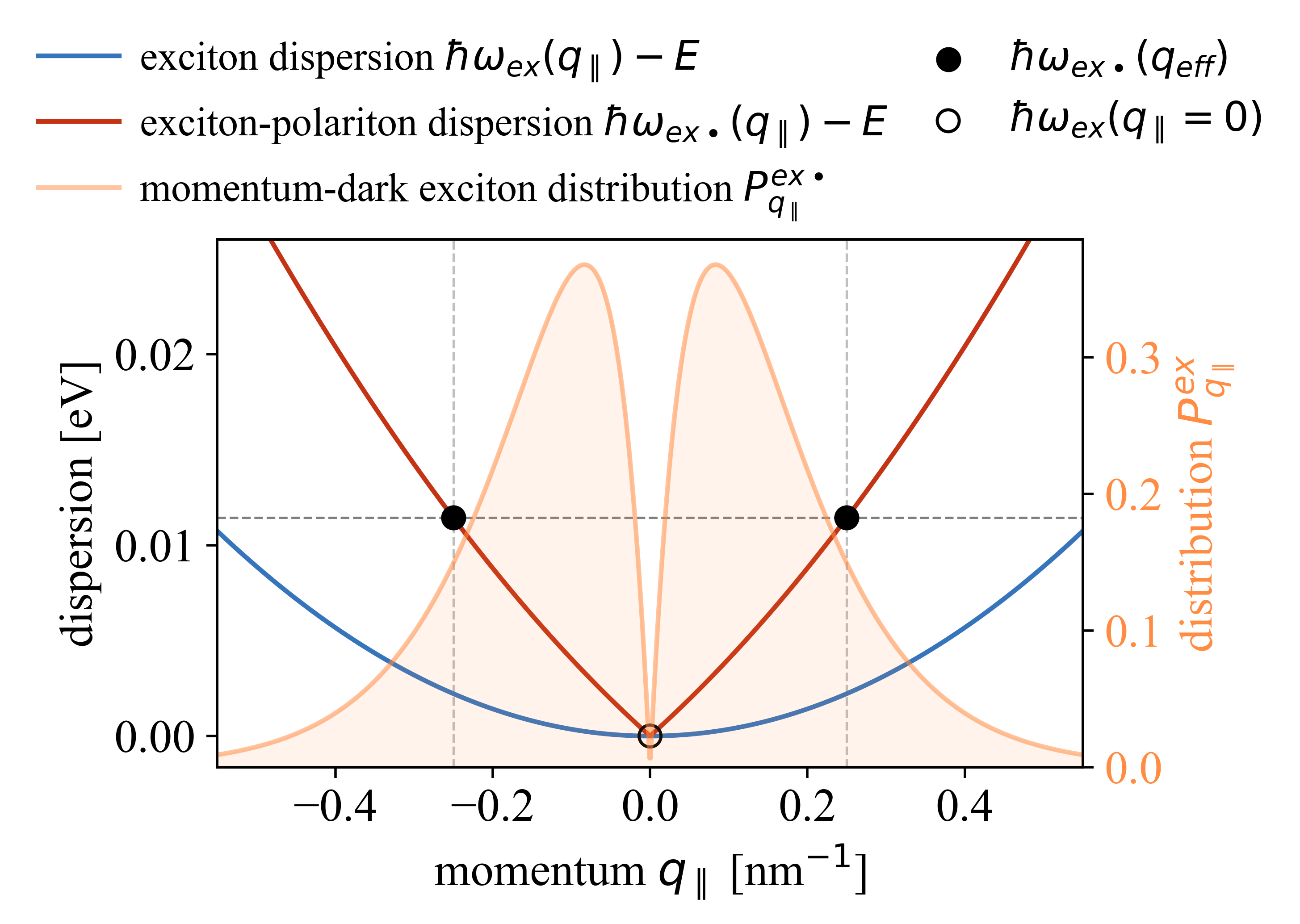}
    \caption{Dispersion relations for exciton respectively exciton-polariton states. The solid lines represent the original exciton ($\hbar\omega_\text{ex}$) and modified exciton-polariton ($\hbar\omega_{\text{ex}\bullet}$) energies as functions of the momentum $q_\parallel$.
    The orange curve shows the exciton distribution $ P^{\text{ex}\bullet}_{q_\parallel} $ with regard to its ordinate axis on the right side.
    Vertical dashed lines indicate the effective exciton momentum $q_{\text{eff}}$. The horizontal dashed line marks the energy $\hbar\omega_{\text{ex}\bullet}(q_{\text{eff}})$ of the effective exciton mode.}
    \label{fig: exciton dispersion}
\end{figure}

Due to the vicinity of plasmon and excitons within a small fraction of the wavelength, we apply the quasi-static approximation \cite{jackson_klassische_2014} to describe their mutual near-field interactions. This limit for the dyadic Green´s function, occurring in Eqs.~(\ref{eq: electric field of TMDC at MNP},\ref{eq: electric field of MNP at TMD}), corresponds to neglecting the frequency compared to the near-field momentum:
\begin{align}
      \varepsilon_\text{out}\frac{\omega^2}{c^2}
      \ll {q_\parallel^2}
      .\label{eq: quasi-static condition}
\end{align}
This assumption is valid for exciton dipole density contributions with high momenta $\mathbf{q}_\parallel$, i.e.,~outside the light-cone, Eq.~\eqref{eq: def out light cone}.
In contrast, the momentum-bright excitons ($q_\parallel = 0$) excited by the incident electric field $\mathbf{E}^0_{\mathbf{q}_\parallel=\mathbf{0}}$ within the light-cone are described with the full frequency dependency separately in Sec.~\ref{sec: bright excitons}.
For momentum-dark excitons, the dyadic Green´s function in quasi-static approximation simplifies to 
\begin{align}
\mathbcal{G}_{\mathbf{q}_\parallel}(z,z^\prime)
&={G}_{\mathbf{q}_\parallel}(z,z^\prime)\frac{{q}_\parallel^2}{\varepsilon_0\varepsilon_\text{out}} \mathbcal{U}_\phi
\end{align}
with $(q_\parallel,\phi)$ the polar coordinates of $\mathbf{q_\parallel}$, as illustrated in~Fig.~\ref{fig: exciton dispersion} and the degenerate, idempotent matrix $\mathbcal{U}_\phi$, Eq.~\eqref{eq: matrix U}.
Using the quasi-static scalar Green's functions
\begin{align}
    {G}_{\mathbf{q}_\parallel}(z,z^\prime)
    &= \frac{-1}{2q_\parallel} e^{-q_\parallel |z-z^\prime|},
\end{align}
we obtain
\begin{align}
\mathbcal{G}_{\mathbf{q}_\parallel}(z_\zplidx,z_\text{ex})
&=\frac{-1}{2\varepsilon_0\varepsilon_\text{out}}\,q_\parallel
e^{-q_\parallel\,\delta z}
\mathbcal{U}_\phi,
\end{align}
where $\delta z = |z_\text{ex}-z_\text{Au}|$ denotes the center to center distance of TMDC and MNP, cp.~Fig.~\ref{fig:system}.
Consequently, the plasmon electric field experienced by the excitons, Eq.~(\ref{eq: electric field of MNP at TMD}), becomes 
\begin{align}
\mathbf{E}^{\Eplidx}_\mathbf{q_\parallel}(z_\text{ex};\omega)
&=
\frac{-1}{2\varepsilon_0\varepsilon_\text{out}}\,q_\parallel
e^{-q_\parallel\,\delta z}
\mathbcal{U}_\phi\cdot \mathbf{p}^\pplidx (\omega). \label{eq: electric field of MNP at TMD 2}
\end{align}
We first insert the plasmons electric field, Eq.~\eqref{eq: electric field of MNP at TMD 2}, in the oscillator equation for momentum-dark excitons, Eq.~\eqref{coupled eq 1: dark exciton},
\begin{align}
&\left( \omega^2 +i\gamma^\text{ex}\omega - \omega_\text{ex}^2(q_\parallel)\right) \mathbf{P}^{\text{ex}\bullet}_\mathbf{q_\parallel}\label{coupled eq 2: exciton}\\
&=\frac{\omega_\text{ex}(q_\parallel)}{\hbar}\,  \frac{\vert\varphi_{0} d\vert^2}{\varepsilon_0\varepsilon_\text{out}}
q_\parallel \mathbcal{U}_\phi\cdot  \left(
{\textbf{P}^{\text{ex}\bullet}_{\mathbf{q_\parallel}}}(\omega)
+
e^{-q_\parallel\,\delta z} 
\mathbf{p}^\pplidx (\omega)
 \right)
.\nonumber
\end{align}

Since our goal is to provide an analytic approach to fit experiments, we apply here a separation ansatz for time and polar momentum dependencies of the excitonic polarization
\begin{align}
\mathbf{P}^{\text{ex}\bullet}_{\mathbf{q}_\parallel}(\omega)
=
P^{\text{ex}\bullet}_{q_\parallel}\,\mathbcal{U}_\phi \cdot \mathbf{P}^{\text{ex}\bullet}(\omega) \label{eq: exciton seperation ansatz}
\end{align}
Inserting this separation ansatz into Eq.~\eqref{coupled eq 2: exciton}, and integrating over the polar angle $\phi$ yields
\begin{align}
&\left( \omega^2 +i\gamma^\text{ex}\omega - \omega_{\text{ex}\bullet}^2(q_\parallel)
\right) P^{\text{ex}\bullet}_{q_\parallel}\, \mathbf{P}^{\text{ex}\bullet}(\omega)\label{eq: exciton HO with momentum dependency}\\
&=\frac{\omega_\text{ex}(q_\parallel)}{\hbar}\, \frac{\vert\varphi_{0} d\vert^2}{\varepsilon_0\varepsilon_\text{out}}
q_\parallel e^{-q_\parallel\,\delta z}
\mathbf{p}^\pplidx (\omega) , \nonumber
\end{align}
where the momentum-dark exciton resonance $\omega_{\text{ex}\bullet}(q_\parallel)$ appears at:
\begin{align}
    \omega_{\text{ex}\bullet}^2 (q_\parallel)  = \omega_{\text{ex}}^2(q_\parallel)+ \frac{\omega_\text{ex}(q_\parallel)}{\hbar}\, 
\frac{\vert\varphi_{0} d\vert^2}{\varepsilon_0\varepsilon_\text{out}} q_\parallel . \label{eq: dark exciton disperion}
\end{align}
Similar to refs.~\onlinecite{qiu_nonanalyticity_2015,glazov_ultrafast_2024}, the modification to the free exciton $\omega_{\text{ex}}(q_\parallel)$, stemming from
the electric field mediated exciton self-interaction, can be
considered a polaritonic correction.
The modified exciton-polariton dispersion, Eq.~\eqref{eq: dark exciton disperion}, is displayed in Fig.~\ref{fig: exciton dispersion}.
With Eq.~\eqref{eq: exciton HO with momentum dependency} we thus encounter a distinct harmonic oscillator equation for the momentum-dark excitons with different momenta corresponding to different energies.
In order to simplify Eq.~\eqref{eq: exciton HO with momentum dependency} for all ${q}_\parallel$ to a single equation,
all these ${q}_\parallel\neq {0}$ excitons are collected into a single, effective exciton mode.
This is accomplished by introducing the exciton resonance frequency $\omega_{\text{ex}\bullet}^2 (q_\parallel)$ at a single effective momentum $q_{\text{eff}}$.
We define $\omega_{\text{ex}\bullet}^2 (q_\parallel) \approx \omega_{\text{ex}\bullet}^2 (q_\text{eff}) \equiv \omega_{\text{ex}\bullet}^2$ (horizontal dotted line in Fig.~\ref{fig: exciton dispersion}).
This approximation is crucial to recover the important physics beyond a fully phenomenological oscillator model, by introducing a collective dark exciton state.
For simplicity, we furthermore assume for the free exciton dispersion, Eq.~\eqref{eq: parabolic exciton dispersion}, $\omega_\text{ex}(q_\parallel) \approx \omega_{\text{ex}\circ}$, where we neglect the quadratic contribution that is small compared to both, $\omega_{\text{ex}\circ}$ and the linear $q_\text{eff}$ dependent correction.
This allows to solve the momentum dependency of Eq.~\eqref{eq: exciton HO with momentum dependency} with the distribution
\begin{align}
    P^{\text{ex}\bullet}_{q_\parallel} =  q_\parallel\,\delta z\,  e^{-q_\parallel\,\delta z}, \label{eq: exciton momentum distribution}
\end{align}
which we depict in Fig.~\ref{fig: exciton dispersion} with parameters provided in table \ref{table: parameters}.
Obviously, the momentum-dark exciton, Eq.~\eqref{eq: exciton momentum distribution}, vanishes at $q_\parallel=0$. \cite{katzer_impact_2023}
We determine the effective exciton momentum $q_{\text{eff}}$ (vertical dotted lines in Fig.~\ref{fig: exciton dispersion}) as the averaged absolute value of the momentum regarding the 2D-momentum distribution, Eq.~\eqref{eq: exciton momentum distribution}:
\begin{align}
    q_{\text{eff}} &=
    \frac{\int_{\mathbb{R}^2}  d^2\mathbf{q}_\parallel \, q_{\parallel}
     P^{\text{ex}}_{q_\parallel}  }
    {\int_{\mathbb{R}^2}  d^2\mathbf{q}_\parallel
       P^{\text{ex}}_{q_\parallel} }
    = \frac{3}{\delta z}
\end{align}
For the frequency dependency of the momentum-dark excitons, we find
\begin{align}
\left( \omega^2 +i\gamma^\text{ex}\omega - \omega_{\text{ex}\bullet}^2 \right) \mathbf{P}^{\text{ex}\bullet}(\omega)
=
\frac{\omega_{\text{ex}\circ}}{\hbar\, \delta z}\, \frac{\vert\varphi_{0} d\vert^2}{\varepsilon_0\varepsilon_\text{out}}
\mathbf{p}^\pplidx (\omega) \label{coupled eq 2.1: exciton}.
\end{align}
In the time domain, Eq.~\eqref{coupled eq 2.1: exciton} corresponds to a harmonic oscillator for the collective momentum-dark exciton mode coupled to the plasmon:
\begin{align}
\left( \partial_t^2 +\gamma^\text{ex}\partial_t + \omega_{\text{ex}\bullet}^2\right) \mathbf{P}^{\text{ex}\bullet}(t)+\frac{\omega_{\text{ex}\circ}}{\hbar \, \delta z}\, \frac{\vert\varphi_{0} d\vert^2}{\varepsilon_0\varepsilon_\text{out}} 
\mathbf{p}^\pplidx (t)
&=0 \label{coupled eq 3: exciton}
\end{align}
The convention of the applied Fourier transform are specified in Appendix \ref{app: Fourier transform}.

\subsection{Coupled oscillator equation for the MNP plasmon\label{sec: coupling of the plasmons}}
The solution for the dark exciton momentum dependency
allows to simplify the expression for their coupling to the plasmon.
From $\frac{q^2_\text{eff}}{\varepsilon_\text{out} }   \gg \frac{ \omega^2}{c^2}$ in the Green's dyadic, Eq.~\eqref{eq:dyadic Green's function}, for MNPs significantly smaller than the wavelength of the incident light, it follows that the action of bright excitons ($q_\parallel \approx 0$) to the MNP plasmon dynamics can be neglected in Eq.~\eqref{eq: electric field of TMDC at MNP}, compared to the near-field stemming from momentum-dark excitons.
The electric field, Eq.~\eqref{eq: electric field of TMDC at MNP}, of the momentum-dark excitons can be treated in quasi-static approximation, similar to the previous section. It simplifies to
\begin{align}
    \mathbf{E}^{\text{ex}}(\mathbf{r}_\rplidx ;\omega)
&= \frac{-1}{2\varepsilon_0\varepsilon_\text{out}} \frac{1}{(2\pi)^2} \int d^2 q_\parallel
\,q_\parallel
e^{-q_\parallel\,\delta z}
\mathbcal{U}_\phi\cdot {\textbf{P}^{\text{ex}\bullet}_{\mathbf{q_\parallel}}}(\omega). \label{eq: electric field of TMD at MNP 2}
\end{align}
Inserting the exciton momentum distribution, Eqs.~(\ref{eq: exciton seperation ansatz},\ref{eq: exciton momentum distribution}), in its emitted electric near-field, Eq.~\eqref{eq: electric field of TMD at MNP 2}, yields
\begin{align}
\mathbf{E}^{\text{ex}}(\mathbf{r}_\rplidx ;\omega)
&= \frac{-1}{4\pi\varepsilon_0\varepsilon_\text{out}}  
\frac{3}{16} \frac{1}{(\delta z)^3}{\textbf{P}^{\text{ex}\bullet}}(\omega). \label{eq: electric field of collective dark-exciton mode at MNP}
\end{align}
Thus, from Eqs.~(\ref{coupled eq 1: plasmon},\ref{eq: electric field of collective dark-exciton mode at MNP}), we find for the plasmon
\begin{align}
    &\left( \omega^2 + i\gamma^\text{pl}(T)  \omega - \omega_{\oplidx}^2 \right) \mathbf{p}^\pplidx (\omega)
    -\omega_{\oplidx}^2 \frac{V_\text{MNP} }{L}
    \frac{3}{64\pi} 
    \frac{1}{{\delta z}^3}
    \mathbf{P}^{\text{ex}\bullet}(\omega)\nonumber\\
    &= -\omega_{\oplidx}^2 \frac{V_\text{MNP} }{L}
     \varepsilon_0\varepsilon_\text{out} \mathbf{E}^0(\mathbf{r}_\rplidx ; \omega) 
    \label{coupled eq 2: plasmon}.
\end{align}
In the time domain, Eq.~\eqref{coupled eq 2: plasmon} yields a harmonic oscillator equation for the MNP plasmon $\mathbf{p}^\pplidx $ coupled to the collective momentum-dark exciton mode $\mathbf{P}^{\text{ex}\bullet}$:

\begin{align}
&\left( \partial_t^2 + \gamma^\text{pl}(T)  \partial_t + \omega_{\oplidx}^2 \right) \mathbf{p}^\pplidx (t)
    +\omega_{\oplidx}^2 \frac{V_\text{MNP} }{L}
    \frac{3}{64\pi} 
    \frac{1}{{\delta z}^3}
    \mathbf{P}^{\text{ex}\bullet}(t)\nonumber\\
    &= \omega_{\oplidx}^2 \frac{V_\text{MNP} }{L}
     \varepsilon_0\varepsilon_\text{out} \mathbf{E}^0(\mathbf{r}_\rplidx ; t) 
    \label{coupled eq 3: plasmon}.
\end{align}

\subsection{Coupled oscillator equation for bright excitons\label{sec: bright excitons}}
In the previous section, we restricted ourselves to the near-field where excitons with high momenta, Eq.~\eqref{eq: def out light cone}, and plasmons interact with evanescent electric fields. However, excitons with small momenta, cp.~Eq.~\eqref{eq: def in light cone}, may also contribute to the optical spectra. In fact, they dominate it for a pristine TMDC since excitons outside the light cone are not excited by far field excitation without the presence of the MNP (or another source that breaks translational invariance and thus provides an evanescent electric field).
To incorporate the bright excitons in the coupled model, we explicitly insert the electric field generated by the MNP, Eq.~\eqref{eq: electric field of MNP at TMD}, evaluated at $\mathbf{q}_\parallel = \mathbf{0}$,
\begin{align}
    \mathbf{E}_{\mathbf{q}_\parallel = \mathbf{0}}^\text{pl} (z_\text{ex},\omega) 
    &=i\omega\,
    \frac{1}{2c}\frac{1}{\varepsilon_0\sqrt{\varepsilon_\text{out}}} \, \mathbf{p}^\pplidx (\omega) \, e^{i \frac{\sqrt{\varepsilon_\text{out}}}{c} \delta z \, \omega},
\end{align}
in the bright exciton dynamics, Eq.~\eqref{coupled eq 1: bright exciton}. 
Transforming into time domain, the frequency prefactor becomes a time derivative and the exponential yields a time delay. We obtain
\begin{align}
&\left( \partial_t^2 +\gamma^\text{ex}_{\circ}\partial_t + \omega_\text{ex}^2\right) \mathbf{P}^{\text{ex}\circ}_\mathbf{q_\parallel=0}(t)\label{eq: 2. bright exciton HO}\\
&=
\frac{2\omega_\text{ex}}{\hbar}\, \vert\varphi_{0} d\vert^2
\left[ 
\mathbf{E}_{\mathbf{q_\parallel=0}}^0 (z_\text{ex};t)
-\frac{1}{2c\sqrt{\varepsilon_\text{out}}\varepsilon_0} \partial_t \mathbf{p}^\pplidx (t-\delta t )
\right] ,\nonumber
\end{align}
where $\delta t =\frac{\sqrt{\varepsilon_\text{out}}}{c}\delta z $ is the propagation time of light from the plasmon position to the TMDC plane.

\subsection{Three Coupled Oscillators Model (3-COM)}
In the previous sections, we found three coupled harmonic oscillators that effectively model the temporal dynamics of the TMDC-MNP hybrid: the plasmon, the collective momentum-dark exciton and the bright exciton oscillator equations. In the following, we summarize these oscillator contributions to the coupled TMDC-MNP dynamics:

\begin{tcolorbox}[customformula, width=0.488\textwidth, title={\vspace{-0.4cm}
\begin{align}\text{Three Coupled Oscillators Model (3-COM)}\label{eq: coupled eqs final}\end{align}}]
\setlength{\jot}{0pt} 
\abovedisplayskip=0pt
\belowdisplayskip=0pt
\begin{align*}
\left( \partial_t^2 + \gamma^\text{pl}(T)  \partial_t + \omega_{\oplidx}^2 \right) \mathbf{p}^\pplidx (t)
    +g^\gplidx
    \mathbf{P}^{\text{ex}\bullet}(t)
    &=f^\fplidx  \mathbf{E}^0(\mathbf{r}_\rplidx ; t) \\
    \\
\left( \partial_t^2 +\gamma^\text{ex}(T)\partial_t + \omega_{\text{ex}\bullet}^2 \right) \mathbf{P}^{\text{ex}\bullet}(t)+ g^\text{ex}_\bullet
\mathbf{p}^\pplidx (t)
&=0 \\
\\
\left( \partial_t^2 +\gamma^\text{ex}_{\circ}(T)\partial_t +\omega_{\text{ex}\circ}^2\right) \mathbf{P}^{\text{ex}\circ}_\mathbf{q_\parallel=0}(t)
\,+\, g^\text{ex}_\circ \partial_t &
\mathbf{p}^\pplidx (t-\delta t)\nonumber\\
= 
f^\text{ex}_\circ &
\mathbf{E}_{\mathbf{q_\parallel=0}}^0  (z_\text{ex};t)
\end{align*}
\end{tcolorbox}

An overview of the occurring parameters is given in table \ref{table: 3-COM parameters}.
Equation \eqref{eq: coupled eqs final} shows that the system is reduced to the coupled dynamics of three oscillators:
The incident electric field directly drives the MNP plasmon and the bright exciton. On the other hand, the momentum-dark exciton couples to the MNP plasmon via the coupling strength $g^{\text{ex}}_\bullet$, and vice versa via $g^{\text{pl}}$. The temporal gradient of the MNP plasmon, with a time delay $\delta t$, corresponding to the light propagation time over the distance $\delta z$, and coupling strength $g^{\text{ex}}_\circ$, affects the bright exciton mode. The impact of the momentum bright exciton on the MNP plasmon is negligible. The first two lines in Eq.~\eqref{eq: coupled eqs final} are similar to the common COM as given in Ref.~\onlinecite{wu_quantum-dot-induced_2010}, but we note that here the momentum-dark exciton frequency $\omega_{\text{ex}\bullet}$ differs from the bright exciton frequency $\omega_{\text{ex}\circ}$ as observed in far-field spectroscopy.
The inclusion of the third oscillator, the bright exciton with its qualitatively different coupling to the MNP plasmon, represents an extension to the common COM.

\begin{table}[t]
\centering
 \caption{Parameter definitions for the 3-COM}
 \renewcommand{\arraystretch}{1.75}
 \begin{tabularx}{\linewidth}{XX}
 \hline\hline
    MNP plasmon & $\mathbf{p}^\text{pl}$\\
    momentum-dark exciton & $\mathbf{P}^{\text{ex}\bullet}$\\
    bright exciton & $\mathbf{P}^{\text{ex}\circ}$\\
    incident electric field & $\mathbf{E}^0$\\
 \hline
    localized surface plasmon frequency & $\omega_{\oplidx} = \omega_\text{pl,bulk} \sqrt{ \frac{ L}{L \varepsilon_b  + (1-L) \varepsilon_\text{out}}}$\\
    dark exciton frequency      & ${\omega_{\text{ex}\bullet}= \sqrt{ \omega_{\text{ex}\circ}^2+\frac{\omega_{\text{ex}\circ}\vert\varphi_{0} d\vert^2 \, q_\text{eff}}{\hbar\,\varepsilon_0\varepsilon_\text{out}}}}$\\
    eff.~dark exciton momentum      & $q_\text{eff} =\frac{3}{\delta z}$\\
         bright exciton frequency      & $\omega_{\text{ex}\circ}$\\\hline
    plasmon coupling constant & $g^\gplidx = \omega_{\oplidx}^2 \frac{V_\text{MNP} }{L}
    \frac{3}{64\pi} 
    \frac{1}{{\delta z}^3}$\\
    dark exciton coupling constant & $g^\text{ex}_{\bullet} = \frac{\omega_{\text{ex}\circ}}{\hbar}\, \vert\varphi_{0} d\vert^2 \frac{1}{\varepsilon_0\varepsilon_\text{out}} \frac{1}{\delta z}$\\
    effective coupling constant & $g_\text{eff}=\sqrt{g^\gplidx g^\text{ex}_{\bullet}}$\\
    bright exciton coupling constant & $g^\text{ex}_{\circ}= \frac{\omega_{\text{ex}\circ}}{\hbar}\, \vert\varphi_{0} d\vert^2
\frac{1}{\sqrt{\varepsilon_\text{out}}\varepsilon_0} \frac{1}{c}$\\
    propagation time of light & $\delta t = \frac{\sqrt{\varepsilon_\text{out}}}{c} \delta z$\\\hline
    plasmon oscillator strength & $f^\fplidx  = \omega_{\oplidx}^2 \frac{V_\text{MNP} }{L}
     \varepsilon_0\varepsilon_\text{out}$\\
    bright exciton oscillator strength & $f^\text{ex}_\circ = \frac{2\omega_{\text{ex}\circ}}{\hbar}\, \vert\varphi_{0} d\vert^2$\\\hline
    plasmon damping constant & $\gamma^\text{pl}(T)$, see~Eqs.~(\ref{eq: drude damping}-\ref{eq: drude damping - elph})\\
    dark exciton damping constant & $\gamma^\text{ex}(T)$, see~Eq.~\eqref{eq: exciton damping definition (1. order)}\\
    bright exciton damping constant & $\gamma^{\text{ex}}_{\circ}(T)$, see~Eqs.~(\ref{eq: exciton damping definition (1. order)},\ref{eq: damping bright exciton})\\\hline\hline
    \label{table: 3-COM parameters}
\end{tabularx}
\vspace{-1cm}
\end{table}

To solve the coupled set of harmonic oscillators, we make a plane wave ansatz for the incident electric field propagating along the $z$-direction
\begin{align}
    \mathbf{E}^0(\mathbf{r},t) = \mathbf{E}^0 \Re \left( e^{ik z-i\omega t} \right)
    = \mathbf{E}^0 \cos \left( k z-\omega t\right) \label{eq: def incident E0}
\end{align}
with $k = \sqrt{\varepsilon_\text{out}}\frac{\omega}{c} $ and the in-plane polarized amplitude $\mathbf{E}^0$.
We solve the coupled set of equations via a Fourier transform, which is straight forward for the complex representation of the electric field and dipole densities. As a last step, we take the real part of the coupled set of linear equations.
The solution for the temporal plasmon, momentum-dark exciton and bright exciton dynamics is given by:
\begin{widetext}
\begin{align}
    \mathbf{p}^\pplidx (t) &=-\Re
    \left( 
    \frac{\Big(\omega^2+i\omega\gamma^\text{ex}-\omega_{\text{ex}\bullet}^2\Big)}{\Big(\omega^2+i\omega\gamma^\text{pl}-\omega_\oplidx^2\Big)
    \Big( \omega^2+i\omega\gamma^\text{ex}-\omega_{\text{ex}\bullet}^2\Big)
    -
    g^\gplidx g^\text{ex}_\bullet}
    e^{ik z_\zplidx-i\omega t}
    \right)
    f^\fplidx \,\mathbf{E}^0
    \label{eq: solution p_pl}\\
\mathbf{P}^{\text{ex}\bullet}(t) &=-\Re
    \left( 
    \frac{g^\text{ex}_\bullet}
    {\Big(\omega^2+i\omega\gamma^\text{pl}-\omega_\oplidx^2\Big)
    \Big( \omega^2+i\omega\gamma^\text{ex}-\omega_{\text{ex}\bullet}^2\Big)
    -
    g^\gplidx g^\text{ex}_\bullet}
    e^{ik z_\zplidx-i\omega t}
    \right)
    f^\fplidx \,\mathbf{E}^0
    \label{eq: solution p_ex dark}\\
\mathbf{P}_{\mathbf{q_\parallel}=0}^{\text{ex}\circ}(t)
    &=
    -\Re \left(\frac{1}{\omega^2+i\gamma^{\text{ex}}_{\circ} \omega -\omega_{\text{ex}\circ}^2} e^{ikz_\text{ex}-i\omega t}\right)
    (2\pi)^2 \delta(\mathbf{q_\parallel}) f_\circ^\text{ex}\mathbf{E}^0
\end{align}
\end{widetext}
For the bright exciton, we neglect the coupling to the plasmon that is weak since $g^\gplidx ,g^\text{ex}_\bullet \gg g^\text{ex}_\circ$. If taken into account, it acts as an additional contribution to the damping $\gamma^{\text{ex}\circ}$ of the bright exciton \cite{greten_strong_2024}.
Considering the definitions in table \ref{table: 3-COM parameters}, we find for a symmetrized form of the coupled oscillators, Eq.~\eqref{eq: coupled eqs final}, the effective coupling strength between momentum-dark exciton and plasmon 
\begin{align}
    g_\text{eff} = \sqrt{g^\gplidx  g^\text{ex}_\bullet}\sim 
    \begin{cases} 
    \sqrt{V_\text{MNP}}\\
    \delta z^{-2}
\end{cases}
\end{align}
that depends on the MNP volume $V_\text{MNP}$ and the exciton-plasmon distance $\delta z$.
Due to the 2d-geometry, the derived distance dependency differs from the coupling of a MNP plasmon to a 0d exciton \cite{todisco_excitonplasmon_2015, yan_optical_2008,swathi_distance_2009}, which typically follows $\delta z^{-3}$.
It is important to note that the dependence of $g_\text{eff}$ on the surrounding permittivity $\varepsilon_\text{out}$, as discussed in Ref.~\onlinecite{goncalves_plasmon-exciton_2018}, cannot be accurately captured within such a simplified oscillator model. For further details, see Appendix \ref{app sec: permittivity dependency}.

The bright $(q_\parallel = 0)$ exciton mode shows no spatial dependence
\begin{align}
    \mathbf{P}^{\text{ex}\circ}(\mathbf{r};t)
    &=
    -\Re \left(\frac{1}{\omega^2+i\gamma^\text{ex}_{\circ} \omega -\omega_{\text{ex}\circ}^2} e^{ikz_\text{ex}-i\omega t}\right) f_\circ^\text{ex}\mathbf{E}^0 \label{eq: solution P_ex bright},
\end{align}
and exists independently on the localized exciton distribution \cite{salzwedel_spatial_2023} stemming from momentum-dark excitons.

\subsection{Observables}

Corresponding to energy balance, the absorbed energy $W_\text{abs}$ is given by the extinction $W_\text{ext}$ and the radiated energy $W_\text{rad}$:
\begin{align}
    W_\text{abs} = W_\text{ext} + W_\text{rad}, \label{eq: energy balance}
\end{align}
where $W_\text{rad}<0$ since it is out-going energy, re-radiated respectively scattered by the system, cp.~Fig.~\ref{fig:system}.
The energy absorbed within the volume $V$ and observation time $\tau$ is given by \cite{jackson_klassische_2014}
\begin{align}
    W_\text{abs} = \int_\tau \!\! dt \int_V \!\! d^3r\, \mathbf{E}(\mathbf{r},t)\cdot  \dot{\mathbf{P}}(\mathbf{r},t). \label{eq: absorption poynting theorem}
\end{align}
A detailed analysis of all arising terms is given in Appendix \ref{sec: Appendix reciprocity principle}. In the following, we focus on the contributions to connect the derived solution for the coupled plasmon-exciton dynamics with measurable optical properties. We provide results for the extinction (Sec.~\ref{sec: extinction}), experimentally accessible via linear transmission and reflection \cite{vadia_magneto-optical_2023,abid_resonant_2016,abid_temperature-dependent_2017,lee_electrical_2017,lee_fano_2015,liu_strong_2016,petric_tuning_2022,wang_coherent_2016,zhang_observation_2023,zhang_steering_2021}, and scattering cross section (Sec.~\ref{sec: scattering cross section}) as measured in dark-field spectroscopy \cite{bisht_collective_2019,cuadra_observation_2018,geisler_single-crystalline_2019,hou_simultaneous_2022,kern_nanoantenna-enhanced_2015,kleemann_strong-coupling_2017,qin_revealing_2020,wen_room-temperature_2017,zheng_manipulating_2017}.

\subsubsection{Extinction \label{sec: extinction}}
The extinction is the energy absorbed and scattered by the system that equals the work done by the incident electric field at the MNP plasmon and the TMDC exciton, cp.~Appendix \ref{sec: Appendix reciprocity principle}.
\begin{align}
    W_\text{ext} = \int_\tau\!\! dt \int_V\!\! d^3r\, \mathbf{E}^0(\mathbf{r},t)\cdot  \dot{\mathbf{P}}(\mathbf{r},t)
\end{align}

Since the incident electric field $\mathbf{E}^0\delta_{\mathbf{q_\parallel},\mathbf{0}}$ acts directly only on originally bright excitons ($q_\parallel=0$) that have been approximated by a constant 2d dipole density and the MNP plasmon within the laser spot with area $A$, only the coupling of these terms to the incident field $\mathbf{E}^0$ contributes.
Therefore, we define with respect to the measured far-field extinction
\begin{align}
    W_\text{ext} = W_\text{ext}^\text{pl}
    +
    W_\text{ext}^{\text{ex}\circ}
\end{align}
with the plasmon and bright exciton extinctions
\begin{align}
    W_\text{ext}^\text{pl} &= \int_\tau\! dt \, \mathbf{E}^0(\mathbf{r}_\rplidx ,t)\cdot  \dot{\mathbf{p}}^\text{pl}(t),\\
    W_\text{ext}^{\text{ex}\circ} &= A \int_\tau \! dt\, \mathbf{E}^0(\mathbf{r}_\parallel,z_\text{ex},t)\cdot  \dot{\mathbf{P}}^{\text{ex}\circ}(\mathbf{r}_\parallel,t).
\end{align}
Since the observation time $\tau$ is significantly longer than a period of the incident light $\tau \gg \frac{2\pi}{\omega}$, we average over the carrier frequency oscillation.
To determine the normalized extinction $w_\text{ext}$, we divide the work done at the hybrid structure $W_\text{ext}$ by the in-coming energy
\begin{align}
    W^0 = A \tau \frac{\sqrt{\varepsilon_\text{out}}}{2} \varepsilon_0 c \vert\mathbf{E}^0\vert^2
\end{align}
on the illuminated area $A$ during time $\tau$.
We obtain the normalized plasmon-exciton extinction
\begin{align}
    w_\text{ext} = \frac{W_\text{ext}^\text{pl} + W_\text{ext}^\text{ex}}{W^0} = w_\text{ext}^\text{pl} + w_\text{ext}^{\text{ex}\circ} \label{eq: normalized work}.
\end{align}
Inserting the solved exciton and plasmon dynamics from the previous section, Eqs.~(\ref{eq: solution p_pl},\ref{eq: solution p_ex dark},\ref{eq: solution P_ex bright}), and the incident electric field, Eq.~\eqref{eq: def incident E0}, we obtain:

\begin{widetext}
\begin{center}
\begin{tcolorbox}[customformula, width=0.75\textwidth, title={\vspace{-0.4cm}
\begin{align}\text{Normalized Extinction}\:\:\: w_\text{ext} = w_\text{ext}^\text{pl} + w_\text{ext}^{\text{ex}\circ}\label{eq: normalized extinction}\end{align}}]
\setlength{\jot}{0pt} 
\abovedisplayskip=0pt
\belowdisplayskip=0pt
\begin{align*}
    w_\text{ext}^\text{pl} &= - \frac{1}{A} \frac{\omega f^\fplidx  }{c\varepsilon_0\sqrt{\varepsilon_\text{out}}} 
    \Im \left(
    \frac{(\omega^2+i\gamma^\text{ex}\omega - \omega_{\text{ex}\bullet}^2)}{(\omega^2+i\gamma^\text{ex}\omega - \omega_{\text{ex}\bullet}^2)(\omega^2+i\gamma^\text{pl}\omega - \omega_\oplidx^2)-g^\gplidx g^\text{ex}_\bullet}
    \right)\\
    w_\text{ext}^{\text{ex}\circ} &= -  \frac{\omega f_\circ^\text{ex}}{c\varepsilon_0\sqrt{\varepsilon_\text{out}}} 
    \Im \left(
    \frac{1}{\omega^2+i\gamma^\text{ex}_{\circ}\omega - \omega_{\text{ex}\circ}^2}
    \right)
\end{align*}
\end{tcolorbox}
\end{center}
\end{widetext}

The MNP plasmon contribution $w_\text{ext}^\text{pl}$ is qualitatively similar to the commonly applied COM \cite{wu_quantum-dot-induced_2010}. However, we obtain additionally the bright exciton contribution $w_\text{ext}^{\text{ex}\circ}$ for TMDC exciton-MNP plasmon hybrids. The exciton and plasmon damping coefficients $\gamma^\text{ex},\,\gamma^\text{pl}$ and $\gamma^\text{ex}_{\circ}$ are temperature dependent, cp.~Eqs.~(\ref{eq: exciton damping definition (1. order)},\ref{eq: damping bright exciton},\ref{eq: drude damping}). Therefore, we obtain qualitatively different spectra for different temperatures.

\subsubsection{Scattering Cross Section \label{sec: scattering cross section}}
The momentum-dark excitons correspond to a localized exciton distribution under the MNP with extensions small compared to the wavelength of the incident light \cite{salzwedel_spatial_2023}. Therefore, the scattering cross section of the hybridized exciton-plasmon system can be modeled based on the scattering cross section of a point dipole \cite{jackson_klassische_2014}
\begin{align}
    S_\text{scat} =
    \frac{1}{6\pi\varepsilon_0^2 } \frac{ \omega^4 }{ c^4}  \left\vert  \alpha^\text{plex}(\omega) \right\vert^2 \label{eq: definition scattering cross section}
\end{align}
with the polarizability $\alpha^\text{plex}$ that in this case is a coupled plasmon-exciton polarizability. It connects the resulting dipole with the incident field and is therefore defined by the comparison of
\begin{align}
    \mathbf{p}^\pplidx (t) = \alpha^\text{plex} (\omega) \,\mathbf{E}^0(t) \label{eq: definition polarizability}
\end{align}
with the solution for the plasmon dynamics in the hybrid system, cp.~Eq.~\eqref{eq: solution p_pl}. Equations~(\ref{eq: definition scattering cross section} and \ref{eq: definition polarizability}) are only defined in the complex representation of electrodynamics where the polarizability is given by
\begin{align}
    \alpha^\text{plex} (\omega)
    =
    \frac{-\Big(\omega^2+i\omega\gamma^\text{ex}-\omega_{\text{ex}\bullet}^2\Big)\, f^\fplidx}{\Big(\omega^2+i\omega\gamma^\text{pl}-\omega_\oplidx^2\Big)
    \Big( \omega^2+i\omega\gamma^\text{ex}-\omega_{\text{ex}\bullet}^2\Big)
    -
    g^\gplidx g^\text{ex}_\bullet}
     .
\end{align}
Thus we find for the scattering cross section:
\begin{widetext}
\begin{center}
\begin{tcolorbox}[customformula, width=0.75\textwidth, title={\vspace{-0.4cm}
\begin{align}\text{Scattering Cross Section}\label{eq: solution scattering cross section}\end{align}}]
\setlength{\jot}{0pt} 
\abovedisplayskip=0pt
\belowdisplayskip=0pt
\begin{align*}
    S_\text{scat} =
    \frac{\left\vert f^\fplidx \right\vert^2}{6\pi\varepsilon_0^2 } \frac{ \omega^4 }{ c^4}  \left\vert
    \frac{\Big(\omega^2+i\omega\gamma^\text{ex}-\omega_{\text{ex}\bullet}^2\Big)}{\Big(\omega^2+i\omega\gamma^\text{pl}-\omega_\oplidx^2\Big)
    \Big( \omega^2+i\omega\gamma^\text{ex}-\omega_{\text{ex}\bullet}^2\Big)
    -
    g^\gplidx g^\text{ex}_\bullet}
    \right\vert^2
\end{align*}
\end{tcolorbox}
\end{center}
\end{widetext}

The translation-invariant, homogeneous distribution of bright excitons in the 2d TMDC does not contribute to the scattering as measured in dark-field spectroscopy but only to reflection and transmission perpendicular to the TMDC plane. Thus, we would not expect the bright excitonic mode to be visible in dark-field measurements.

\section{Results \label{sec: results}}

To put the results of the three coupled oscillator model (3-COM) into the context of the known model of two coupled oscillators we first discuss two undamped, harmonic oscillators in strict resonance: In this ideal case, the estimated mode-splitting $\Omega_g$ follows the coupling constant via \cite{torma_strong_2014}
\begin{align}
    \Omega_g = \hbar \frac{g_\text{eff}}{\omega_\oplidx}.
\end{align}
Whether the mode splitting manifests as a peak splitting in optical spectra depends on its ratio to the spectral linewidths, that is, the damping coefficients of the individual oscillators. The hybrid system is assigned to weak coupling if the mode splitting is smaller than the mean damping coefficient, $\Omega_g < (\gamma^\text{ex}+\gamma^\text{pl})/2$, and to strong coupling if it is larger, $\Omega_g > (\gamma^\text{ex}+\gamma^\text{pl})/2$.

\begin{figure}[b]
\vspace{-0.5cm}
\centering
    \includegraphics[width=\linewidth]{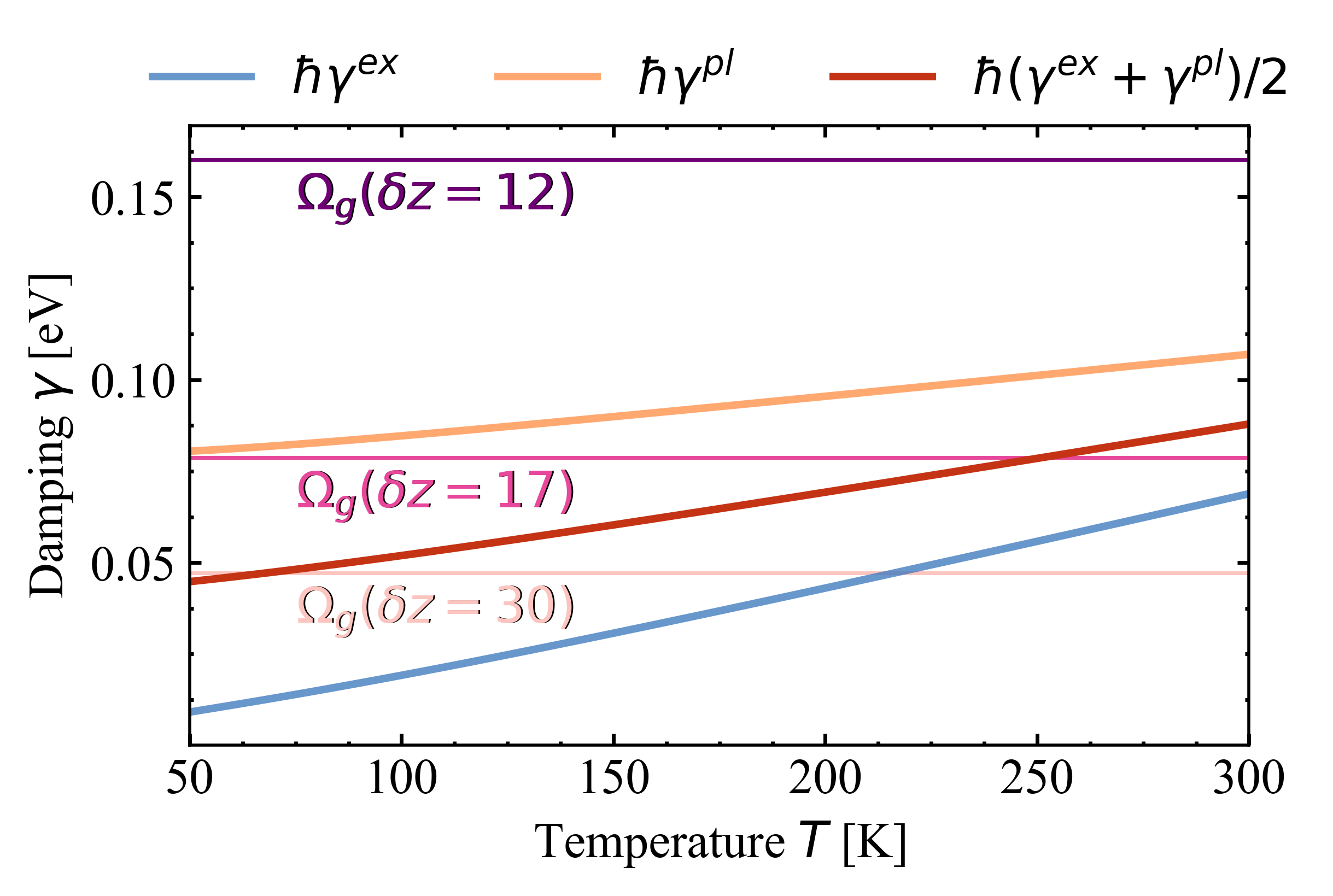}
    \caption{Comparison of mode splittings $\Omega_g$ (thin horizontal lines) for different exciton-plasmon distances $\delta z$ with the temperature-dependent damping of momentum-dark exciton $\gamma^\text{ex}$ (blue), plasmon $\gamma^\text{pl}(T)$ (orange) and with their mean value (red).}
    \label{fig: damping vs coupling}
\end{figure}

\begin{figure*}[t]
\centering
    \includegraphics[width=\linewidth]{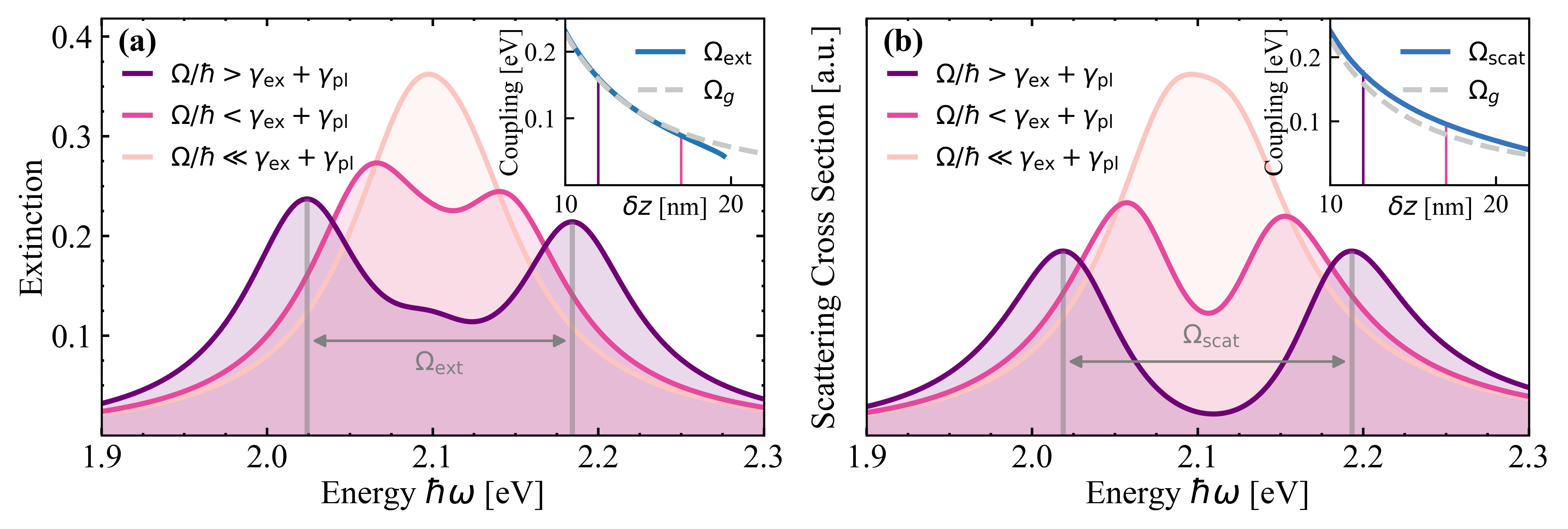}
    \caption{(a) Extinction / (b) Scattering cross section for different distances $\delta z =\{ 12, 17, 30\}\,$nm between the TMDC and the MNP at room temperature $T=293\,$K. The figure shows the distance dependence of exciton-plasmon coupling and its influence on the line shape, that is, a transition from weak coupling (large distances) to strong coupling (small distances). Insets: Comparison between the distance dependency of the coupling constant ($\sim \Omega_g$) and the peak splitting extracted from the respective spectrum (a) $ \Omega_\text{ext}$ and (b) $ \Omega_\text{scat}$. For weak coupling strengths the peak splitting vanishes which is extinction at $\delta z \gtrsim 20\,$nm.}
    \label{fig: z Scan}
\end{figure*}

The parameters used in this entire section are provided in table \ref{table: parameters}.
The temperature-dependent damping coefficients are depicted in Fig.~\ref{fig: damping vs coupling}, where the mean value is compared with the estimated mode splitting $\Omega_g$ for different distances $\delta z\in\{ 12,17,30\}\,$nm between the TMDC layer and MNP center, cp.~Fig.~\ref{fig:system}.
For small distances $(\delta z = 12\,\text{nm})$, the estimated mode splitting $\Omega_g$ remains in the strong coupling regime across the entire temperature range $T \in [50, 300]$, lying well above the mean damping coefficient. At larger distances, the system shifts to the weak coupling regime.

We now turn to the resulting optical spectra.
Figures \ref{fig: z Scan}a and \ref{fig: z Scan}b show extinction, Eqs.~(\ref{eq: normalized work},\ref{eq: normalized extinction}), and scattering, Eq.~\eqref{eq: solution scattering cross section}, at room temperature for the different distances $\delta z$. 
These spectra show the transition from weak coupling at larger distances, e.g., $\delta z = 30\, \text{nm}$, to strong coupling at smaller distances, comparable to the extensions of the MNP (e.g., $\delta z = 12\, \text{nm} \gtrsim r_z$).

The insets in Figs.~\ref{fig: z Scan}a and \ref{fig: z Scan}b compare the distance dependence of the estimated mode splitting $\Omega_g$ with the peak splitting extracted from the respective spectrum ($\Omega_\text{ext/scat}$).
For extinction, Fig.~\ref{fig: z Scan}a, there is a very good agreement between $\Omega_\text{ext}$ and $\Omega_g$ in the strong coupling regime.
However, in scattering, Fig.~\ref{fig: z Scan}b, $\Omega_\text{scat}$ deviates from $\Omega_g$. The reason for this deviation is the $\omega^4$ prefactor of the dipole scattering efficiency, cp.~Eq.~\eqref{eq: solution scattering cross section}.

\begin{table}[h!]
\centering
 \caption{Material parameters for TMDC and MNP}
 \begin{tabularx}{\linewidth}{XXXc}
 \hline\hline
    Parameter & Value &Unit& Reference \\
    \hline
    $d$      & $0.25$       & $e$ nm                                       & \cite{xiao_coupled_2012}\\
    $M$           &  $6.1$        &  eV fs$^2$ nm$^{-2}$     & \cite{kormanyos_k_2015}\\
    $c_1$  &   $0.182$       & meV\,K$^{-1}$        & \cite{selig_excitonic_2016}\\
    $c_2$  &  $15.6$   $31.2$     & meV         & \cite{selig_excitonic_2016}\\
    $\Omega$  &  $30$         & meV        & \cite{selig_excitonic_2016}\\
   $\varphi_0$       &     $0.46$  & nm$^{-1}$ & {$^{\text{a}}$}\\ \hline
   $\varepsilon_\infty$  &  $1.53$ & & \cite{etchegoin_analytic_2006}\\
   $ \omega_\text{pl,bulk}$ & $12.99$ &fs$^{-1}$ & \cite{etchegoin_analytic_2006}\\
    $ \omega_1$ &  $4.02$ & fs$^{-1}$ & \cite{etchegoin_analytic_2006}\\
    $ \omega_2$ &  $5.69$ & fs$^{-1}$ & \cite{etchegoin_analytic_2006}\\
    $ \Gamma_1$ &  $0.82$ & fs$^{-1}$ & \cite{etchegoin_analytic_2006}\\
    $ \Gamma_2$ &  $2.00$ & fs$^{-1}$ & \cite{etchegoin_analytic_2006}\\
   $A_1$ & $0.94$ & & \cite{etchegoin_analytic_2006}\\
   $A_2$ & $1.36$ & & \cite{etchegoin_analytic_2006}\\
   $\phi_1$ & $-\pi/4$ & & \cite{etchegoin_analytic_2006}\\
   $\phi_2$ & $-\pi/4$ & & \cite{etchegoin_analytic_2006}\\
   $b$ & $0.6329$ & eV$^{-1}$ & \cite{liu_reduced_2009}\\
   $\gamma_0$ & $0.0219$ & eV & \cite{liu_reduced_2009}\\
   $\Theta$ & $185$ & K & \cite{liu_reduced_2009}\\\hline
    $c$             &    $299.7925$    &  nm fs$^{-1}$                       &  \\
    $\varepsilon_0$             &    $0.05526308$ &  e$^2$ eV$^{-1}$ nm$^{-1}$                       &  \\
    $\hbar$             &    $0.658212196$    &  eV\,fs                       &  \\
    $k_B$             &    $0.0861745$    &  meV\,K$^{-1}$                       &  \\
    \hline
   $r_{x}$ &$30$& nm&\\
   $r_{y}$ &$30$& nm&\\
   $r_{z}$ &$10$& nm&\\
    $\delta z$ &$12$& nm&\\
    $\varepsilon_\text{out}$ &$3$& & \\
    $A$ &$16000$& nm$^2$ & \\
   \hline\hline \label{table: parameters}
\end{tabularx}
\vspace{-0.25cm}
\flushleft{
\footnotesize{
        {{$^{\text{a}}$}} appears as the solution of the Wannier equation, similar to Refs.~\cite{rytova_screened_1967,keldysh_coulomb_1979}, for the chosen dielectric environment. 
    }
}
\end{table}

\begin{figure*}[t]
\centering
    \includegraphics[width=\linewidth]{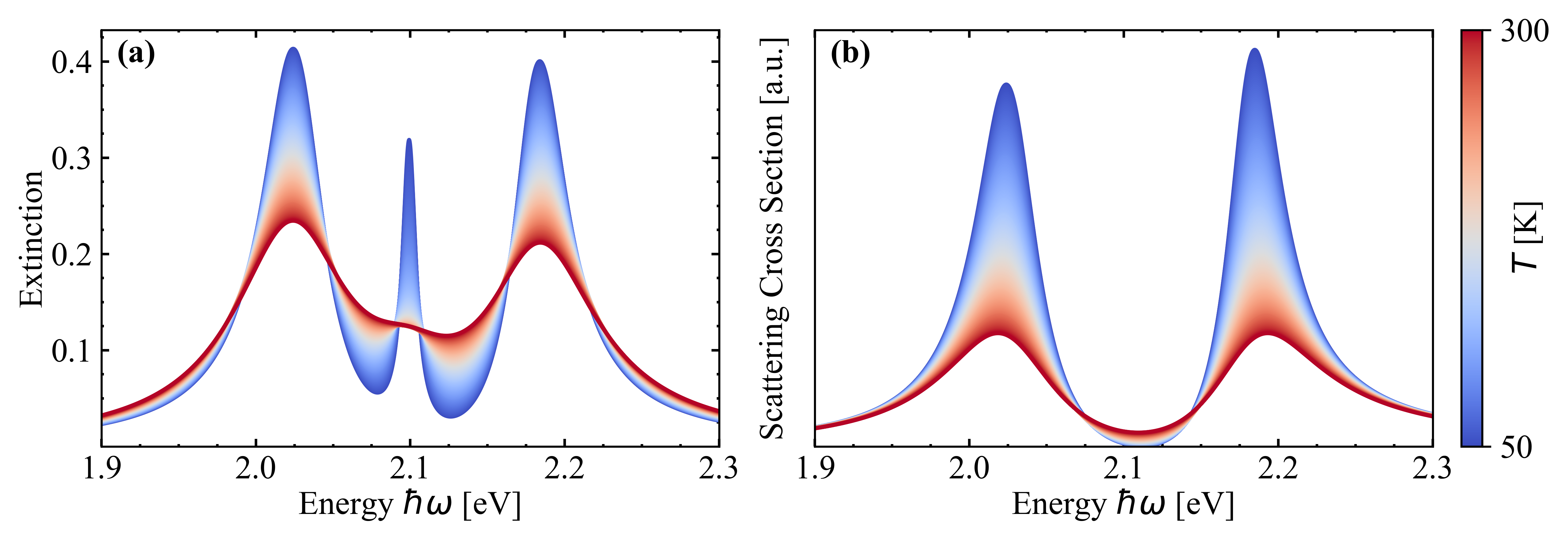}
    \caption{Spectra for different temperatures T. (a) Extinction: at high temperatures, the spectrum resembles almost a typical strong coupling spectrum of two coupled oscillators, with the higher energy peak showing a slightly reduced amplitude due to the detuning between the plasmon and the collective momentum-dark exciton mode. The additional bright exciton peak becomes more pronounced at lower temperatures. (b) Scattering cross section: It includes only the scattering at the MNP plasmon, influenced by the coupling to the TMDC excitons, displaying typical strong coupling characteristics.} \label{fig: T Scan}
\end{figure*}

Furthermore, the extinction spectrum in fig.~\ref{fig: z Scan}a differs in another aspect from the ideal case:
For two perfectly resonant oscillators, the amplitude of the hybridized modes in extinction spectra would be exactly equal. In our case, despite setting $\omega_\oplidx=\omega_{\text{ex}\circ}$, we observe a slight asymmetry between the two peaks. The asymmetry arises because the plasmon is resonant with the bright exciton mode but not with the momentum-dark excitons, which are strongly coupled but lie at higher energies. Since the light-matter interaction of the momentum-dark excitons is mediated by the plasmon, the lower energy mode, which has more plasmonic character when the plasmon resonance frequency is lower than the momentum-dark exciton, is more pronounced.
In fact, the 3-COM underestimates this asymmetry by summarizing the continuum of momentum-dark energy states in one effective mode. A semi-analytical study without this approximation, necessitating a numerical evaluation, is conducted in Ref.~\onlinecite{greten_strong_2024}.
For the scattering cross section, an asymmetry between the modes is characteristic due to the $\omega^4$ dependence even in the perfectly resonant case, making the high-energy mode more pronounced. This effect counteracts the asymmetry caused by the effective detuning of the hybridized oscillators.

For extinction, the strong coupling case in Fig.~\ref{fig: z Scan}a hints to the emergence of a third, central peak. 
To better resolve this feature, Figs.~\ref{fig: T Scan}a and \ref{fig: T Scan}b show the extinction and scattering spectra, respectively, for varying temperatures. In the model, the temperature influences both the exciton and plasmon damping coefficients, leading to qualitatively different spectra.
At room temperature, the spectra exhibit the typical strong coupling characteristics of two coupled oscillators.
As the temperature decreases, an additional central peak becomes visible in the extinction because the reduced damping allows for a clear observation of the bright exciton mode. In contrast, the delocalized bright excitons have no contribution to the scattering cross section, according to the model in Sec.~\ref{sec: scattering cross section}, and thus can not be observed in dark-field measurements.

\begin{figure}[t]
\centering
    \includegraphics[width=\linewidth]{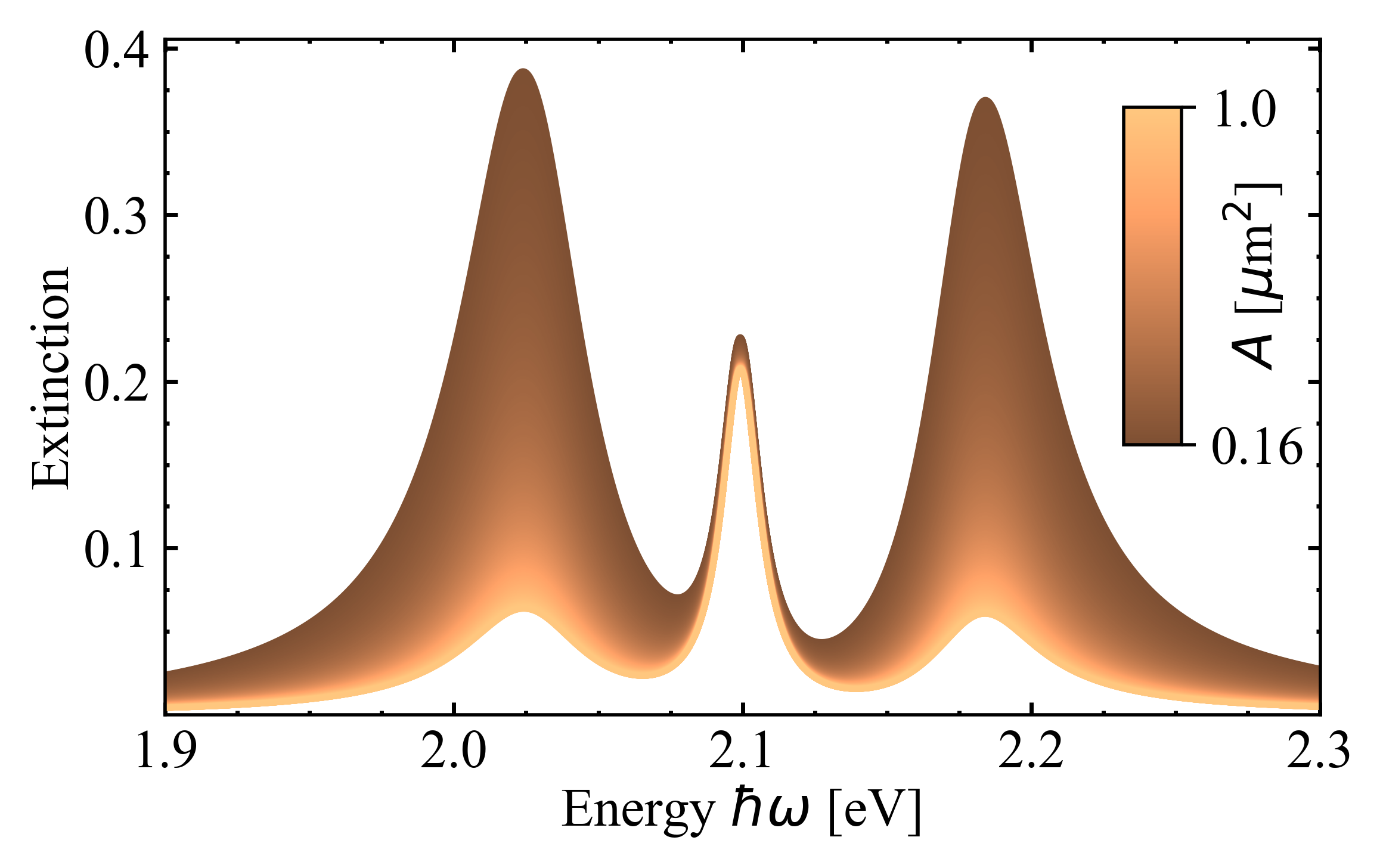}
    \caption{Extinction for different sizes of the illuminated area $A$ per MNP at liquid nitrogen temperature $T=77\,$K. The larger the illuminated area of the TMDC per MNP, the more pronounced is the middle, bright exciton in relation to the hybridized plasmon - dark exciton modes.}
    \label{fig: A Scan absorption}
\end{figure}

According to Eq.~\eqref{eq: normalized work}, the ratio of the hybridized plasmon-dark exciton modes and the bright exciton contribution to the extinction depends on both temperature $T$ and the illuminated area $A$ of the TMDC per MNP.
In samples with multiple MNPs (assuming no interactions between them), this parameter corresponds to the inverse of the MNP density.
The assumption of negligible interactions between MNPs is valid for randomly distributed MNPs and a certain range of distances $(\sim 3 r_x)$ \cite{mueller_microscopic_2018,greten_strong_2024,barros_plasmon-polaritons_2021}.
We evaluate the extinction for varying $A\in [0.16,1]\,\mu$m at liquid nitrogen temperature in Fig.~\ref{fig: A Scan absorption}. For a laser spot size $A=0.16\,\mu$m$^2$ right at the diffraction limit with a single MNP on the TMDC, the spectrum is dominated by hybridized plasmon - dark exciton modes. With increasing $A$, the bright exciton mode becomes more pronounced.

\section{Conclusion}
\label{sec:conclusion}

We developed a modified three coupled oscillator model (3-COM) to describe experimentally accessible observables, such as extinction and scattering, in hybrid systems composed of metal nanoparticles (MNPs) and two-dimensional delocalized excitons, e.g., in transition metal dichalcogenides (TMDCs).
For the occurring coupling parameters we provide explicit expressions depending on distance of MNP and TMDC, MNP number density, MNP volume and shape as well as material properties.
Most importantly, the 3-COM incorporates key extensions to the conventional COM by accounting for the spatial delocalization of two-dimensional excitons, distinguishing between bright and momentum-dark exciton modes, and their interactions with MNP plasmons.

We have demonstrated that while bright excitons remain weakly coupled, momentum-dark excitons can exhibit strong coupling with MNP plasmons but lie at higher energies. This results in the appearance of a distinct third (bright exciton) peak in the optical spectra \cite{vadia_magneto-optical_2023} and an asymmetry between the hybridized plasmon-dark exciton modes \cite{cuadra_observation_2018,geisler_single-crystalline_2019,kleemann_strong-coupling_2017,liu_strong_2016,petric_tuning_2022,qin_revealing_2020,wen_room-temperature_2017}, consistent with experimental observations.

The visibility of the bright exciton mode in the hybrid spectrum is determined by temperature and MNP density, though it is not observable in dark-field spectra, which only detect scattered light.

While the simplified 3-COM offers significant advantages, particularly in its ability to fit experimental data, it involves certain approximations. 
The assumption that the MNP plasmon fully covers the absorption (e.g., Drude susceptibility) excludes possible contributions from inner metal bands, limiting the model to a certain frequency range and reducing its ability to fully account for the impact of surrounding permittivity. 
On the exciton side, the effective treatment of momentum-dark excitons as a single mode simplifies the model but underestimates the resulting asymmetry in the hybridized modes \cite{greten_strong_2024,salzwedel_spatial_2023}, as a continuum of excitonic energy states is compressed into one representative energy level. This simplification also leads to an overestimation of the coupling constant, that, however, is no limitation for the application of this model for fitting experimental spectra.

In summary, our analytical approach bridges the gap between traditional phenomenological models \cite{wu_quantum-dot-induced_2010,torma_strong_2014,limonov_fano_2017,joe_classical_2006} and rigorous microscopic theory \cite{gurrieri_dynamics_2024,salzwedel_spatial_2023,greten_strong_2024,denning_quantum_2022} for TMDC-MNP hybrid systems. It offers a straightforward and physically grounded framework for fitting and interpreting experimental results in nanoscale optical systems using Eqs.~(\ref{eq: normalized extinction} and \ref{eq: solution scattering cross section}), which together with the derived oscillator model, Eq.~\eqref{eq: coupled eqs final}, constitute the main result of our work.


\section{Acknowledgments}
We would like to thank Jonas Grumm, Robert Fuchs, Joris Sturm, Frank Jahnke and David Greten for their insightful comments on various topics related to this work.
Further, we acknowledge the assistance of generative AI (ChatGPT) in improving the language clarity of this manuscript.



\appendix

\section{Fourier transform\label{app: Fourier transform}}
The Fourier transform regarding the in-plane spatial coordinates $\mathbf{r}_\parallel = (x,y)^T$ and its inverse for an arbitrary scalar or vector function $\mathbf{f}$ are given by:
\begin{align}
\mathbf{f}(\mathbf{r})&= \frac{1}{(2\pi)^2} \int d^2 \mathbf{q}_\parallel\, e^{i\mathbf{q}_\parallel\cdot\mathbf{r}_\parallel} \mathbf{f}(\mathbf{q}_\parallel,z)\\
\mathbf{f}(\mathbf{q}_\parallel,z)&=\int d^2\mathbf{r}_\parallel\, e^{-i\mathbf{q}_\parallel\cdot\mathbf{r}_\parallel} \mathbf{f}(\mathbf{r})
\end{align}
The Fourier transform with respect to time is defined with opposite sign in the exponential compared to the spatial transformation.
\begin{align}
\mathbf{f}(t)&=\frac{1}{2\pi}\int d\omega\, e^{-i\omega t} f(\omega)\\
\mathbf{f}(\omega)&=\int dt\, e^{i\omega t} f(t)
\end{align}

\section{Reduction from 3 to 2 dimensions}
\label{sec: Appendix to 2 dim}

In the main text, we state that it is sufficient to treat the in-plane directions of the MNP dipole and the TMDC dipole density, dropping their $z$-entries. From this follows that also the description of the resulting electric fields and so the effective Green's dyadic simplifies significantly.
Here, we provide a proof for this statement.

In the developed theory, we need the Green's dyadic twice: 1.~To determine the electric field emitted by the TMDC excitons, observed at the position of the MNP
\begin{align}
\mathbf{E}^{\text{ex}}(\mathbf{r}_\rplidx ;\omega)
&= \frac{1}{(2\pi)^2} \int d^2 q_\parallel
\mathbcal{G}_{\mathbf{q_\parallel}}(z_\zplidx,z_\text{ex};\omega)\cdot {\mathbf{P}^{\text{ex}\,}_{\mathbf{q_\parallel}}}(\omega) \label{eq Appendix: electric field of TMDC at MNP}
\end{align}
and 2.~vice versa,
the electric field emitted by the MNP plasmon, observed in the TMDC monolayer
\begin{align}
\mathbf{E}^{\Eplidx}_\mathbf{q_\parallel}(z_\text{ex};\omega)= \mathbcal{G}_{\mathbf{q_\parallel}}(z_\text{ex},z_\zplidx;\omega)\cdot \mathbf{p}^\pplidx (\omega). \label{eq Appendix: electric field of MNP at TMD}
\end{align}
The full $3\times 3$ Green's dyadic (in-plane Fourier transformed) is given by
\begin{align}
&\mathbcal{G}_{{\mathbf{q_\parallel}}}(z,z',\omega) = \nonumber \\
 & \ \ \begin{pmatrix}
-\frac{\omega^2}{{\varepsilon}_0c^2}\mathbb{1} + \frac{{\mathbf{q_\parallel}}\otimes\,{\mathbf{q_\parallel}}}{{\varepsilon}_0\varepsilon_\text{out}}
 & \frac{i{\mathbf{q_\parallel}}} {{\varepsilon}_0\varepsilon_\text{out}}\partial_{z'} \\ & \\
\frac{i\mathbf{q_\parallel}^T}{{\varepsilon}_0\varepsilon_\text{out}}\partial_{z'} &
-\frac{\omega^2}{{\varepsilon}_0c^2} - \frac{1}{{\varepsilon}_0\varepsilon_\text{out}}\partial_{z'}^2 
\end{pmatrix}
G_{{q_\parallel}} (z,z',\omega). \label{eq Appendix:dyadic Green's function}
\end{align}
We now step by step drop components of the Green's dyadic, Eq.~\eqref{eq Appendix:dyadic Green's function}, that are not necessary to represent Eqs.~(\ref{eq Appendix: electric field of TMDC at MNP} and \ref{eq Appendix: electric field of MNP at TMD}). 

Due to the in-plane orientation of the TMDC exciton dipole moment, we first know that the exciton dipole density has a negligible $z$-entry.
\begin{align}
    {\mathbf{P}^{\text{ex}\,}_{\mathbf{q_\parallel}}}(\omega)
    &=
    \begin{pmatrix}
        {{P}^{\text{ex}\,}_{\mathbf{q_\parallel}x}}
        \\
        {{P}^{\text{ex}\,}_{\mathbf{q_\parallel}y}}\\
        0
    \end{pmatrix}
\end{align}
By representing ${\mathbf{P}^{\text{ex}\,}_{\mathbf{q_\parallel}}}$ as a $2$-dimensional vector in Eq.~\eqref{eq Appendix: electric field of TMDC at MNP}, the Green's dyadic there effectively reduces to $3\times 2$.
Second, to determine the exciton dynamics we need $\mathbf{d}^\xi\cdot \mathbf{E}^{\Eplidx}_\mathbf{q_\parallel}$ which does not depend on the $z$-component of the electric field. Therefore, there is no need to calculate the $z$-entry of $\mathbf{E}^{\Eplidx}_\mathbf{q_\parallel}$, allowing to reduce the dimensions of the Green's dyadic in Eq.~\eqref{eq Appendix: electric field of MNP at TMD} to $2\times 3$.

To get rid of the remaining $z$-entries, we start with the ansatz, that also the MNP dipole has a vanishing $z$-entry
\begin{align}
    \mathbf{p}^\pplidx (\omega)
    &=
    \begin{pmatrix}
        {{p}^\pplidx _{x}}
        \\
        {{p}^\pplidx _{y}}\\
        0
    \end{pmatrix}  \label{eq Appendix: MNP dipole inplane },
\end{align}
which is not yet proven to self-consistently solve the exciton-plasmon coupling dynamics.
However, if this assumption is valid, the dimensions of Eqs.~\eqref{eq Appendix: electric field of TMDC at MNP} and \eqref{eq Appendix: electric field of MNP at TMD} may be reduced to $2\times 2$ (similar arguments as above).
Thus, we finally have to show that from an initially in-plane oriented MNP dipole, excited by the in-plane oriented incident electric field, the coupling to the TMDC acts in such a way, that its back-action does not excite the $z$-component of the MNP dipole.
Following the definition of the polarizability
\begin{equation}
\mathbf{p}^\pplidx (\omega) = \boldsymbol{\alpha}(\omega)\cdot \mathbf{E}(\mathbf{r}_\rplidx ,\omega) \label{app eq: ansatz pz=0}
\end{equation}
with $\boldsymbol{\alpha}$ being diagonal in a Cartesian basis, it follows that ${{p}^\pplidx _{z}}= 0$ if and only if ${{E}_{z}}(\mathbf{r}_\rplidx ;\omega) = {{E}^0_{z}}(\mathbf{r}_\rplidx ;\omega)+{{E}^\text{ex}_{z}}(\mathbf{r}_\rplidx ;\omega)= 0$. 
Since we only consider an incoming electric field $\mathbf{E}^0$ with in-plane orientation, ${{E}^0_{z}}(\mathbf{r}_\rplidx ;\omega)= 0$ is valid.
The $z$-entry of the electric field emitted by the excitons can be determined using Eq.~\eqref{eq Appendix: electric field of TMDC at MNP}. The $z$-row of the remained $3\times 2$ Green's dyadic, Eq.~\eqref{eq Appendix:dyadic Green's function}, is odd in ${q}_{\parallel x}$ and ${q}_{\parallel y}$ since the scalar Green's function only depends on ${q}_{\parallel} = \vert \mathbf{q}_{\parallel} \vert$.
Thus, the relevant $\mathbf{q}_\parallel$ integral vanishes and ${{E}^\text{ex}_{z}}(\mathbf{r}_\rplidx ;\omega)=0$ is valid, if all entries of the exciton dipole density ${\mathbf{P}^{\text{ex}\,}_{\mathbf{q_\parallel}}}$ are even, which in fact is fulfilled if ${{p}^\pplidx _{z}}=0$.
In summary, the ansatz, Eq.~\eqref{app eq: ansatz pz=0}, self-consistently solves the coupled light-matter respectively exciton-plasmon dynamics, and may be described effectively using only $2$d dipole densities without any further approximation.


\section{Energy balance \label{sec: Appendix reciprocity principle}}
In this Appendix, we evaluate the energy balance for the MNP-TMDC hybrid under illumination with the incident electric field $\mathbf{E}^0$ to find an expression for the extinction:
The total absorbed energy is
\begin{align}
    W_\text{abs} = \int_\tau\!\! dt \int_V \!\! d^3r\, \mathbf{E}(\mathbf{r},t)\cdot  \dot{\mathbf{P}}(\mathbf{r},t) \label{app eq: absorption poynting theorem}. 
\end{align}
In our case, the full electric field is given by
\begin{align}
   \mathbf{E}(\mathbf{r};t) = \mathbf{E}^{\text{ex}\circ}(\mathbf{r};t)+
   \mathbf{E}^{\text{ex}\bullet}(\mathbf{r};t)+\mathbf{E}^\Eplidx(\mathbf{r};t)+\mathbf{E}^0(\mathbf{r};t) \label{app eq: full electric field}
\end{align}
and the full dipole density by
\begin{align}
   \mathbf{P}(\mathbf{r};t) = \mathbf{P}^{\text{ex}\circ} (\mathbf{r};t) + \mathbf{P}^{\text{ex}\bullet} (\mathbf{r};t)+\mathbf{P}^{\text{pl}} (\mathbf{r};t) \label{app eq: full dipole density}
\end{align}
with the TMDC dipole density in thin film approximation and the MNP reduced to a point dipole:
\begin{align}
   \mathbf{P}^{\text{ex}\circ} (\mathbf{r};t) = \mathbf{P}^{\text{ex}\circ} (\mathbf{r}_\parallel;t) \delta(z-z_\text{ex})\label{app eq: bright exciton P 2d vs 3d}\\
    \mathbf{P}^{\text{ex}\bullet} (\mathbf{r};t) = \mathbf{P}^{\text{ex}\bullet}(\mathbf{r}_\parallel;t) \delta(z-z_\text{ex})\label{app eq: dark exciton P 2d vs 3d}\\
   \mathbf{P}^{\text{pl}} (\mathbf{r};t) = \mathbf{p}^\pplidx (t) \delta(\mathbf{r}-\mathbf{r}^\text{pl}) \label{app eq: plasmon dipole density}
\end{align}

Taking into account Eqs.~(\ref{app eq: full electric field},\ref{app eq: full dipole density}), the total absorbed energy, Eq.~\eqref{app eq: absorption poynting theorem}, can be divided into the following contributions
\begin{align}
    W_\text{abs} = &\,
    W^{\text{ex}\bullet\leftrightarrow\text{pl}}
    +W^{\text{ex}\circ\leftrightarrow\text{pl}}
    +W^{\text{ex}\circ\leftrightarrow\text{ex}\bullet}\label{app eq: all energy contributions}\\
    &+ W^{\text{ex}\circ\text{,rad}}
    + W^{\text{ex}\bullet\text{,rad}}
    + W^{\text{pl,rad}}
    +W^0. \nonumber
\end{align}
The individual terms are:

\begin{enumerate}
\item The work done by the plasmon on the dark exciton distribution plus vice versa
\begin{align}
    W^{\text{ex}\bullet\leftrightarrow\text{pl}} &= \int_\tau\!\! dt \int_V\!\! d^3r\, \mathbf{E}^\Eplidx(\mathbf{r},t)\cdot  \partial_t
    \mathbf{P}^{\text{ex}\bullet}(\mathbf{r};t)\\
    &+\int_\tau\!\! dt \int_V\!\! d^3r\, \mathbf{E}^{\text{ex}\bullet}(\mathbf{r},t)\cdot  \partial_t
    \mathbf{P}^{\text{pl}}(\mathbf{r},t) = 0. \nonumber
\end{align}
It vanishes due to energy conservation, since the work done by the plasmon on the dark exciton distribution is in turn received by the excitons. We prove the relation in Sec.~\ref{Appendix sec: reciprocity principle}.

\item We disregard the energy exchange between bright exciton and plasmon here by stating
\begin{align}
    W^{\text{ex}\circ\leftrightarrow\text{pl}} &= \int_\tau\!\! dt \int_V\!\! d^3r\, \mathbf{E}^\Eplidx(\mathbf{r},t)\cdot  \partial_t
    \mathbf{P}^{\text{ex}\circ}(\mathbf{r};t)\\
    &+\int_\tau\!\! dt \int_V\!\! d^3r\, \mathbf{E}^{\text{ex}\circ}(\mathbf{r},t)\cdot  \partial_t
    \mathbf{P}^{\text{pl}}(\mathbf{r},t) = 0,\nonumber
\end{align}
to stay consistent with the approximations of the main paper, where we neglect the corresponding minor interaction.

\item 
The energy exchange between bright and dark excitons vanishes, 
\begin{align}
    W^{\text{ex}\circ\leftrightarrow\text{ex}\bullet} &= \int_\tau\!\! dt \int_V\!\! d^3r\, \mathbf{E}^{\text{ex}\bullet}(\mathbf{r},t)\cdot  \partial_t
    \mathbf{P}^{\text{ex}\circ}(\mathbf{r};t)\\
    &+ \int_\tau\!\! dt \int_V\!\! d^3r\,\mathbf{E}^{\text{ex}\circ}(\mathbf{r},t)\cdot  \partial_t
    \mathbf{P}^{\text{ex}\bullet}(\mathbf{r},t) = 0,\nonumber
\end{align}
since they do not interact (see Sec.~\ref{app: darx-bright excitons energy exchange}).

\item The out-going (radiated), respectively scattered energy
\begin{align}
    W_\text{rad} = 
     W^{\text{ex}\bullet\text{,rad}}
    +W^{\text{ex}\circ\text{,rad}}
    + W^{\text{pl,rad}}
\end{align}
with the contributions
\begin{align}
    W^{\text{ex}\circ\text{,rad}} &= \int_\tau\!\! dt \int_V\!\! d^3r\, \mathbf{E}^{\text{ex}\circ}
    (\mathbf{r},t)\cdot  \partial_t
    \mathbf{P}^{\text{ex}\circ}(\mathbf{r};t),\\
    W^{\text{ex}\bullet\text{,rad}} &= \int_\tau\!\! dt \int_V\!\! d^3r\, \mathbf{E}^{\text{ex}\bullet}
    (\mathbf{r},t)\cdot  \partial_t
    \mathbf{P}^{\text{ex}\bullet}(\mathbf{r};t),\\
    W^{\text{pl}\text{,rad}} &= \int_\tau\!\! dt \int_V\!\! d^3r\, \mathbf{E}^{\Eplidx}
    (\mathbf{r},t)\cdot  \partial_t
    \mathbf{P}^{\text{pl}}(\mathbf{r};t),
\end{align}
is dealt with in Sec.~\ref{app: reradiated energy}.

\item The work done by the incident electric field on the dipole density distribution
\begin{align}
    W^0 =  \int_\tau\!\! dt \int_V\!\! d^3r\, \mathbf{E}^0(\mathbf{r},t)\cdot  \dot{\mathbf{P}}(\mathbf{r},t).
\end{align}
is dealt with in Sec.~\ref{app: detailed calculation extinction}.
Since the remaining terms of the energy balance, Eq.~\eqref{app eq: all energy contributions}, are
\begin{align}
    W_\text{abs} = W^0 + W_{\text{rad}}
\end{align}
and extinction is by definition the sum of absorption and scattered/re-radiated energy
\begin{align}
    W_\text{ext} = W_\text{abs} - W_\text{rad}
\end{align}
with $W_\text{rad}<0$,
we find that the work done by the incident electric field on the full dipole density corresponds to extinction
\begin{align}
    W^0 = W_\text{ext}.
\end{align}
Since the scattering / radiated energy $W_\text{rad}$ is included, the quantity $W^0 = W_\text{ext}$ is typically larger than the true absorption $W_\text{abs}$.
\end{enumerate}

In the explicit evaluation of all terms for the hybrid TMDC-MNP geometry, we have to take care, that the approximations are consistent with the approximations made in the main paper.

\subsection{$W^{\text{ex}\bullet\leftrightarrow\text{pl}}$: plasmon - dark exciton quasi-static energy exchange}
\label{Appendix sec: reciprocity principle}

We consider two exemplary dipole density distributions $\mathbf{P}^1(\mathbf{r};t)$ and $\mathbf{P}^2(\mathbf{r};t)$ that interact via the electric near-field in quasi-static approximation and may correspond to the MNP plasmon $\mathbf{P}^\pplidx (\mathbf{r},t)$ and TMDC dark exciton $\mathbf{P}^{\text{ex}\bullet}(\mathbf{r},t)$ dipole density.
The electro-magnetic fields $\mathbf{E}^l,\, \mathbf{B}^l$ with $l\in \{ 1,2\}$ generated by these distributions individually obey Maxwell's equations. Thus
\begin{align}
    \nabla \times \mathbf{B}^l
    = \mu_0 \partial_t \mathbf{P}^l
    + \mu_0 \varepsilon_0 \varepsilon \partial_t \mathbf{E}^l \label{Appendix eq: Maxwell 4}.
\end{align}
We multiply Eq.~\eqref{Appendix eq: Maxwell 4} from the left with $\mathbf{E}^{\bar{l}}$ where $l\neq \bar{l}\in \{1,2\}$ and add the two resulting equations.
\begin{align}
   \mathbf{E}^2 & \cdot \left( \nabla \times \mathbf{B}^1\right)
   +
   \mathbf{E}^1 \cdot \left( \nabla \times \mathbf{B}^2 \right) =  \\
    &\mu_0 \mathbf{E}^2 \cdot  \partial_t \mathbf{P}^1
    +\mu_0 \mathbf{E}^1 \cdot  \partial_t \mathbf{P}^2\nonumber \\
    &+ \mu_0 \varepsilon_0 \varepsilon \mathbf{E}^2 \cdot  \partial_t \mathbf{E}^1
    + \mu_0 \varepsilon_0 \varepsilon \mathbf{E}^1 \cdot  \partial_t \mathbf{E}^2. \nonumber
\end{align}
Integrating over a sufficiently large volume $V$ and an oscillation period $\tau = \frac{2\pi}{\omega}$ yields
\begin{align}
   &\int_\tau dt \int_V d^3r \left( \mathbf{E}^2 \cdot \left( \nabla \times \mathbf{B}^1\right)
   +
   \mathbf{E}^1 \cdot \left( \nabla \times \mathbf{B}^2 \right) \right) \label{Appendix eq: proof reciprocity principle 1} \\
    &=
    \int_\tau dt \int_V d^3r
    \left( \mu_0 \mathbf{E}^2 \cdot  \partial_t \mathbf{P}^1
    +\mu_0 \mathbf{E}^1 \cdot  \partial_t \mathbf{P}^2 \right)\nonumber\\
    &+
    \int_\tau dt \int_V d^3r
    \left(\mu_0 \varepsilon_0 \varepsilon \mathbf{E}^2 \cdot  \partial_t \mathbf{E}^1
    + \mu_0 \varepsilon_0 \varepsilon \mathbf{E}^1 \cdot  \partial_t \mathbf{E}^2 \right) \nonumber.
\end{align}
First, we treat the first line of Eq.~\eqref{Appendix eq: proof reciprocity principle 1}. We find
\begin{align}
    &\mathbf{E}^2 \cdot \left( \nabla \times \mathbf{B}^1 \right)
    +
    \mathbf{E}^1 \cdot \left( \nabla \times \mathbf{B}^2 \right)\\
    & =  \nabla \cdot \left( \mathbf{E}^2\times \mathbf{B}^1 + \mathbf{E}^1\times \mathbf{B}^2 \right),\nonumber
\end{align}
where we used that the curl $\nabla\times \mathbf{E}^l$ vanishes, since these electric fields have been treated in the quasi-static limit in the main paper and therefore are pure gradient fields. Following Gauss' theorem, the divergence allows to transform the integral over the volume $V$ to an integral over the boundary area $\partial V$.
We choose $V$ to be large, such that $\partial V$ is sufficiently far away and the integral vanishes, since the electric near-fields $\mathbf{E}^1$ and $\mathbf{E}^2$ decrease exponentially with the distance. Therefore, we find
\begin{align}
   &\int_\tau dt \int_V d^3r \left( \mathbf{E}^2 \cdot \left( \nabla \times \mathbf{B}^1\right)
   +
   \mathbf{E}^1 \cdot \left( \nabla \times \mathbf{B}^2 \right) \right) = 0.
\end{align}
Next, we turn our attention to the last line of Eq.~\eqref{Appendix eq: proof reciprocity principle 1}. In our case, the electric fields are monochromatic with the frequency $\omega$. The time dependency of the electric fields is therefore generally given by
\begin{align}
    \mathbf{E}^l \sim \cos(\omega t - \varphi^l) 
\end{align}
with the individual phase shift $\varphi^l$. We further define the abbreviation $\varphi^1-\varphi^2 = \delta \varphi$.
It follows that
\begin{align}
    &\int_\tau dt 
    \left(\mathbf{E}^2 \cdot  \partial_t \mathbf{E}^1
    + \mathbf{E}^1 \cdot \partial_t \mathbf{E}^2 \right) \\
    &\sim
    \int_\tau dt 
    \left(\cos(\omega t)\sin(\omega t + \delta\varphi) 
    + \cos(\omega t+ \delta\varphi) \sin(\omega t)  \right) \nonumber\\
    &=
    \int_\tau dt 
    \sin(2\omega t+ \delta\varphi)  \nonumber\\ 
    &=
    0 \nonumber
\end{align}
vanishes since $\tau = \frac{2\pi}{\omega}$.
Therefore, from Eq.~\eqref{Appendix eq: proof reciprocity principle 1}, it follows
\begin{align}
    W^{1\leftrightarrow 2} &= \int_\tau\!\! dt \int_V\!\! d^3r
    \left( \mu_0 \mathbf{E}^2 \cdot  \partial_t \mathbf{P}^1
    +\mu_0 \mathbf{E}^1 \cdot  \partial_t \mathbf{P}^2 \right) =0, 
\end{align}
All assumptions made in this derivation for the distributions $1$ and $2$ are valid for the MNP plasmon and TMDC momentum dark exciton distribution.
Therefore, we deduce
\begin{align}
    W^{\text{ex}\bullet\leftrightarrow \text{pl}} =0.
\end{align}

\subsection{$W^{\text{ex}\circ\leftrightarrow\text{ex}\bullet}$ : bright exciton - dark exciton energy exchange  \label{app: darx-bright excitons energy exchange}}
The energy exchange between bright and dark excitons is given by
\begin{align}
    W^{\text{ex}\circ\leftrightarrow\text{ex}\bullet} &= \int_\tau\!\! dt \int_A \!\! d^2r_\parallel\, \mathbf{E}^{\text{ex}\bullet}(\mathbf{r}_\parallel,z_\text{ex},t)\cdot  \partial_t
    \mathbf{P}^{\text{ex}\circ}(\mathbf{r}_\parallel;t)\\
    &+ \int_\tau\!\! dt \int_A \!\! d^2r_\parallel\, \mathbf{E}^{\text{ex}\circ}(\mathbf{r}_\parallel,z_\text{ex},t)\cdot  \partial_t
    \mathbf{P}^{\text{ex}\bullet}(\mathbf{r}_\parallel;t),\nonumber
\end{align}
where the 3d integral over the volume $V$ reduces to 2d over the area $A$ due to the thin film approximation for the exciton dipole densities, cp.~Eqs.~(\ref{app eq: bright exciton P 2d vs 3d},\ref{app eq: dark exciton P 2d vs 3d}).
The bright exciton dipole density respectively its emitted electric field (stemming from the $q_\parallel=0$ mode) are constant regarding the in-plane coordinate and therefore regarding the spatial integral over $\mathbf{r}_\parallel$.
For the dark exciton dipole density or its emitted electric field, the spatial integral can then be interpreted as a Fourier-transform, evaluated at $\mathbf{q}_\parallel=0$. This yields
\begin{align}
    W^{\text{ex}\circ\leftrightarrow\text{ex}\bullet} &= \int_\tau\!\! dt \, \mathbf{E}_\mathbf{q_\parallel = 0}^{\text{ex}\bullet}(z_\text{ex},t)\cdot  \partial_t
    \mathbf{P}^{\text{ex}\circ}(\mathbf{r}_\parallel;t)\\
    &+ \int_\tau\!\! dt \, \mathbf{E}^{\text{ex}\circ}(\mathbf{r}_\parallel,z_\text{ex},t)\cdot  \partial_t
    \mathbf{P}_\mathbf{q_\parallel = 0}^{\text{ex}\bullet}(t).\nonumber
\end{align}
As discussed in the main paper, the dark exciton distribution and its emitted electric field vanish at  $\mathbf{q}_\parallel=0$. It follows
\begin{align}
    W^{\text{ex}\circ\leftrightarrow\text{ex}\bullet} =0,
\end{align}
which is physically intuitive since we found no electric field mediated interactions at all between bright and momentum dark excitons in the main paper.

\subsection{$W_\text{rad}$ : radiated/scattered energy \label{app: reradiated energy}}
The light emitted by the hybrid system carries the energy
\begin{align}
    W_\text{rad} = 
     W^{\text{ex}\bullet}_\text{rad}
    +W^{\text{ex}\circ}_\text{rad}
    + W^{\text{pl}}_\text{rad}.
\end{align}
The individual energy radiatively lost by a dipole density distribution $\mathbf{P}^{j}$, $j\in \{\text{ex}\bullet, \text{ex}\circ,  \text{pl} \}$, is given by
\begin{align}
    W^{j}_\text{rad} = \int_\tau dt \int_V d^3r\, \mathbf{E}^j(\mathbf{r},t)\cdot  \partial_t \mathbf{P}^{j}(\mathbf{r},t).
\end{align}
For $j= \text{ex}\bullet$, we insert the emitted electric field via the dyadic Green's function in quasi-static approximation
\begin{align}
    \mathbf{E}^{\text{ex}\bullet}(\mathbf{r},t) = \int_V d^3r^\prime\, \mathcal{G}(\mathbf{r}-\mathbf{r}^\prime)\cdot \mathbf{P}^{\text{ex}\bullet}(\mathbf{r}^\prime,t)
\end{align}
and the separation ansatz, which corresponds to the separation ansatz in the main paper but in real space,
\begin{align}
\mathbf{P}^{\text{ex}\bullet}(\mathbf{r},t)
=
P^{\text{ex}\bullet}(\mathbf{r})\, P^{\text{ex}\bullet}(t)\, \mathbf{e}_P \label{app eq: exciton seperation ansatz}.
\end{align}
with the unit vector $\mathbf{e}_P$.
Thus, we find:
\begin{align}
    W^{\text{ex}\bullet}_\text{rad} =&\, 
    \int_\tau dt   P^{\text{ex}\bullet}(t) \, \partial_t P^{\text{ex}\bullet}(t)\\
     \int_V&\!\! d^3r\,
    \left( \int_V\!\! d^3r^\prime\, \mathcal{G}(\mathbf{r}-\mathbf{r}^\prime)\cdot P^{\text{ex}\bullet}(\mathbf{r}^\prime)\,  \mathbf{e}_P\right)
    \cdot
    P^{\text{ex}\bullet}(\mathbf{r})\,   \mathbf{e}_P
    \nonumber
\end{align}
The time dependence is 
$P^{\text{ex}\bullet}(t) \sim \cos(\omega t + \varphi^{\text{ex}\bullet})$. Therefore, the integral over the time period $\tau$ vanishes and we find
\begin{align}
    W^{\text{ex}\bullet}_\text{rad} = 0,
\end{align}
which is reasonable since the dark excitons do not directly emit radiation.

The bright exciton dipole distribution and its emitted electric field are constant regarding the in-plane coordinate $\mathbf{r}_\parallel$. With Eq.~\eqref{app eq: bright exciton P 2d vs 3d}, we find 
\begin{align}
    W^{\text{ex}\circ}_\text{rad} = A \int_\tau dt\, \mathbf{E}^{\text{ex}\circ}(\mathbf{r}_\parallel, z_\text{ex},t)\cdot  \partial_t \mathbf{P}^{\text{ex}\circ}(\mathbf{r}_\parallel,t). \label{app eq: bright exciton radiated energy 1}
\end{align}
Utilizing the Green's dyadic, Eq.~\eqref{eq: Green's function, omega dependent} in the main paper, evaluated at $\mathbf{q}_\parallel = \mathbf{0}$ and the solution for $\mathbf{P}^{\text{ex}\circ}$, Eq.~\eqref{eq: solution P_ex bright} in the main paper, yields
\begin{align}
    \mathbf{E}^{\text{ex}\circ}(\mathbf{r}_\parallel, z_\text{ex},t)
    =
    \frac{\omega}{c} \frac{1}{2\varepsilon_0\sqrt{\varepsilon_\text{out}}} \mathbf{P}^{\text{ex}\circ}
    \left(\mathbf{r}_\parallel,t-\frac{\pi}{2\omega}\right).
\end{align}
We insert the solution for $\mathbf{P}^{\text{ex}\circ}$, Eq.~\eqref{eq: solution P_ex bright}, as well as the parameter $f^\text{ex}_\circ$ from~table \ref{table: 3-COM parameters} in the main paper and evaluate the normalized exciton re-radiated energy, Eq.~\eqref{app eq: bright exciton radiated energy 1}, around the resonance ($\omega_{\text{ex}\circ} + \omega \approx 2 \omega $). We find
\begin{align}
    w^{\text{ex}\circ}_\text{rad} =   \frac{W^{\text{ex}\circ}_\text{rad}}{W^0}
    \approx -
    \frac{1}{2} \frac{(\gamma^{\text{ex}}_\circ- \gamma^\text{ex}) (\gamma^{\text{ex}}_\circ- \gamma^\text{ex})}{\vert \omega -\omega_{\text{ex}\circ}+\frac{i}{2}\gamma^{\text{ex}}_\circ \vert ^2}, \label{app eq: normalized exciton radiated energy}
\end{align}
where $(\gamma^{\text{ex}}_\circ- \gamma^\text{ex})$ indicates solely the radiative losses. $w^{\text{ex}\circ}_\text{rad}$ is negative since it accounts for out-going energy.

For the MNP plasmon, the concept of splitting the extinction into absorption and re-radiated energy is similar. However, the explicit evaluation of the re-radiated energy $W^\text{pl}_\text{rad}$, that for the plasmon corresponds to scattering, gives rise to unphysical infinities.
The reason is that the point-dipole approximation breaks down for the energy of the MNP plasmon in its own emitted electric field.
However, we still observe the common proportionality of extinction to the square of the polarizability.
A derivation of the scattering cross section of a small particle for the example of a dielectric sphere considering the geometry by a given polarizability is provided in Ref.~\onlinecite{jackson_klassische_2014}.
For the scope of this manuscript, we limit ourselves to the explicit evaluation of the radiative emission of the 2D excitons. This sufficiently serves as a proof of concept for assigning the both terms to their respective physical interpretations as re-radiated/scattered energy for bright exciton and MNP plasmon.

\subsection{$W^0 = W_\text{ext}$ : extinction \label{app: detailed calculation extinction}}
In the previous sections, we identified the radiated/scattered energy and proved that the contribution of the energy exchange to the energy balance vanish.
Therefore, the extinction that is the only remaining term, equals the energy that the incident electric field looses on the system in the first place.
\begin{align}
    W_\text{ext} = W^0 = \int_\tau dt \int_V d^3r\, \mathbf{E}^0(\mathbf{r},t)\cdot  \dot{\mathbf{P}}(\mathbf{r},t)
\end{align}
Here, it consists of three contributions
\begin{align}
    W_\text{ext}^{\text{pl}} = \int_\tau dt \int_V d^3r\, \mathbf{E}^0(\mathbf{r},t)\cdot  \partial_t \mathbf{P}^{\text{pl}}(\mathbf{r},t),\\
     W_\text{ext}^{\text{ex}\circ} = \int_\tau dt \int_V d^3r\, \mathbf{E}^0(\mathbf{r},t)\cdot  \partial_t \mathbf{P}^{\text{ex}\circ}(\mathbf{r},t),\\
     W_\text{ext}^{\text{ex}\bullet} = \int_\tau dt \int_V d^3r\, \mathbf{E}^0(\mathbf{r},t)\cdot  \partial_t \mathbf{P}^{\text{ex}\bullet}(\mathbf{r},t).
\end{align}
First, we show that last term, $W_\text{ext}^{\text{ex}\bullet}$ vanishes since the incident electric field does not directly act on the dark-exciton dipole density. The thin film approximation, Eq.~\eqref{app eq: dark exciton P 2d vs 3d}, yields
\begin{align}
    W_\text{ext}^{\text{ex}\bullet} = \int_\tau dt \int_A d^2r_\parallel\, \mathbf{E}^0(\mathbf{r}_\parallel, z_\text{ex},t)\cdot  \partial_t \mathbf{P}^{\text{ex}\bullet}(\mathbf{r}_\parallel,t).
\end{align}
The considered incident electric field $\mathbf{E}^0$ is constant regarding the in-plane coordinate $\mathbf{r}_\parallel$. Therefore the $\mathbf{r}_\parallel$-integral can be treated as a Fourier transform of the ex$\bullet$-dipole density, evaluated at $\mathbf{q}_\parallel=0$. Since $\mathbf{P}^{\text{ex}\bullet}_\mathbf{q\parallel}(t) = \mathbf{0}$, it follows \begin{align}
    W_\text{ext}^{\text{ex}\bullet} = 0.
\end{align}
The full extinction is therefore given by the plasmon and bright exciton contribution
\begin{align}
    W_\text{ext} = W_\text{ext}^{\text{pl}}  + W_\text{ext}^{\text{ex}\circ}.
\end{align}
It is evaluated in the main paper, Eq.~\eqref{eq: normalized extinction}.
To highlight the relation between extinction, absorption and radiative losses, we convert the result for the normalized exciton extinction $w_\text{ext}^\text{ex}$, Eq.~\eqref{eq: normalized extinction} from the main paper, further. Around the resonance, where $\omega_{\text{ex}\circ} + \omega \approx 2 \omega $, we find
\begin{align}
    w_\text{ext}^\text{ex} 
    \approx
    \frac{1}{2} \frac{(\gamma^{\text{ex}}_\circ- \gamma^\text{ex}) \gamma^{\text{ex}}_\circ}{\vert \omega -\omega_{\text{ex}\circ}+\frac{i}{2}\gamma^{\text{ex}}_\circ \vert ^2} \label{app eq: normalized exciton extinction - resonant}.
\end{align}
$(\gamma^{\text{ex}}_\circ- \gamma^\text{ex})$ indicates solely the radiative losses. 
Combining the normalized exciton radiated energy, Eq.~\eqref{app eq: normalized exciton radiated energy}, and extinction, Eq.~\eqref{app eq: normalized exciton extinction - resonant}, allows to determine the exciton absorption
\begin{align}
    w_\text{abs}^\text{ex} = w^{\text{ex}}_\text{ext}+ w^{\text{ex}\circ}_\text{rad} 
    \approx
    \frac{1}{2} \frac{(\gamma^{\text{ex}}_\circ- \gamma^\text{ex})  \gamma^\text{ex}}{\vert \omega -\omega_{\text{ex}\circ}+\frac{i}{2}\gamma^{\text{ex}}_\circ \vert ^2}. \label{app eq: normalized exciton absorption}
\end{align}
This result confirms the allocation of the individual terms to extinction, radiated energy and absorption, since it matches the absorption that can be derived from the Elliot susceptibility for excitons, cp.~e.g.~Ref.~\onlinecite{kira_semiconductor_2012}.

\section{Dependency on surrounding permittivity $\varepsilon_\text{out}$}
\label{app sec: permittivity dependency}

To describe the MNP as a single effective harmonic oscillator, we had to neglect all absorption effects aside from the plasmonic absorption (intraband transitions in the metal conduction band).
In particular, the absorption caused by interband transitions in the background permittivity $\varepsilon_b$ is omitted that is the approximation $\Im (\varepsilon_b)\approx 0$.

\begin{figure}[b]
\centering
\includegraphics[width=\linewidth]{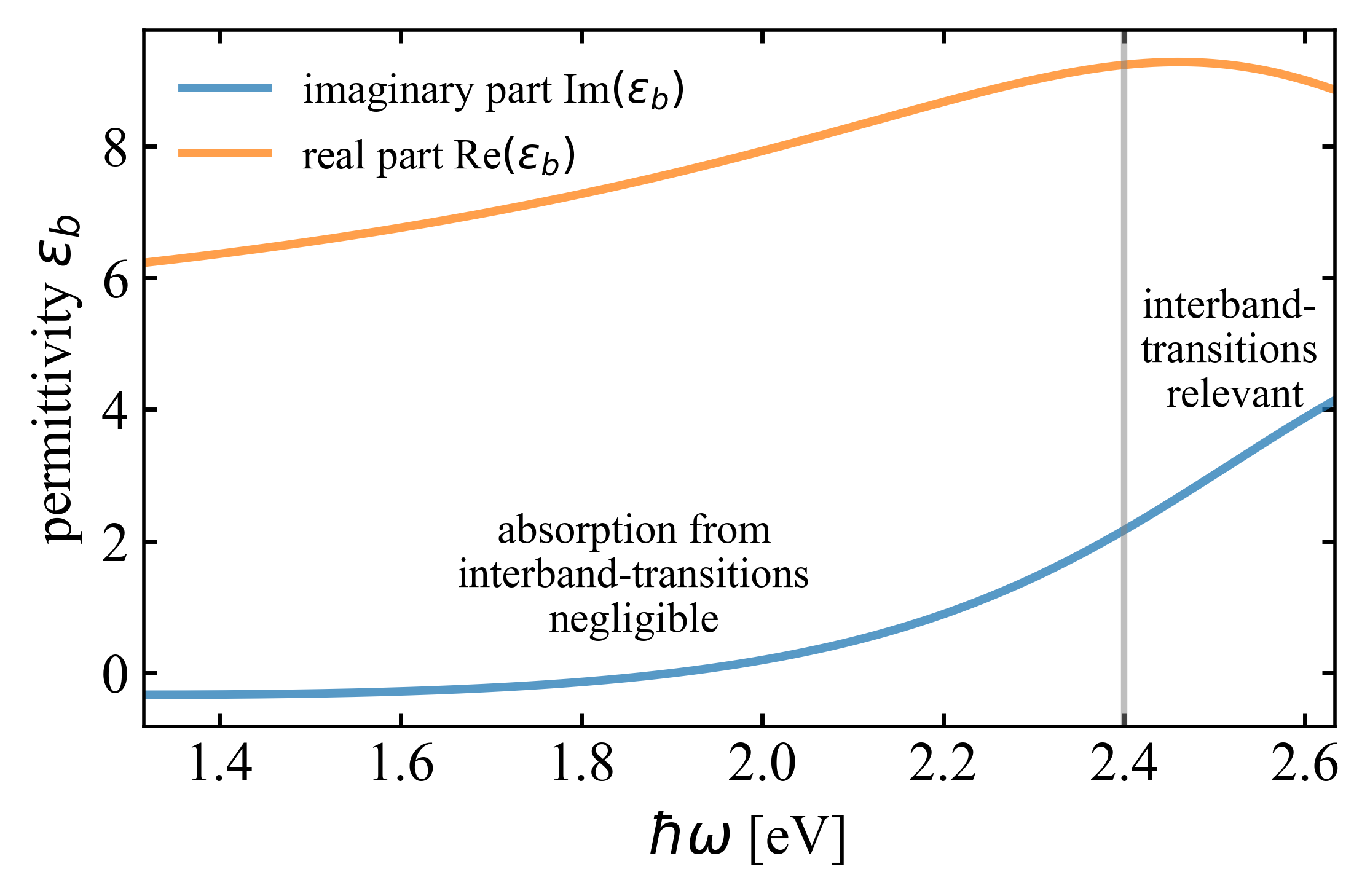}
    \caption{real and imaginary part of non-Drude permittivity $\varepsilon_b$ of the MNP. The imaginary part corresponds to the absorption due to interband transitions. The real part incorporates screening due to inner shells. The approximation $\Im (\varepsilon_b) \approx 0$ (corresponds to neglecting absorption from interband-transitions) is only valid for $\hbar\omega <2.4\,$eV (vertical line)}
    \label{app fig: eps_b}
\end{figure}

Fig.~\ref{app fig: eps_b} illustrates the real and imaginary parts of the background permittivity (non-Drude) of gold, $\varepsilon_b(\omega)$, as defined in Eq.~\eqref{eq: MNP permittivity}. The real part $\Re (\varepsilon_b(\omega))$ (screening) dominates over the imaginary part $\Im (\varepsilon_b(\omega))$ (absorption) in the range $\hbar\omega < 2.4\,$eV, which limits the validity of our model to these energies, where interband transitions are negligible.

\begin{figure*}[t]
\centering
    \includegraphics[width=\linewidth]{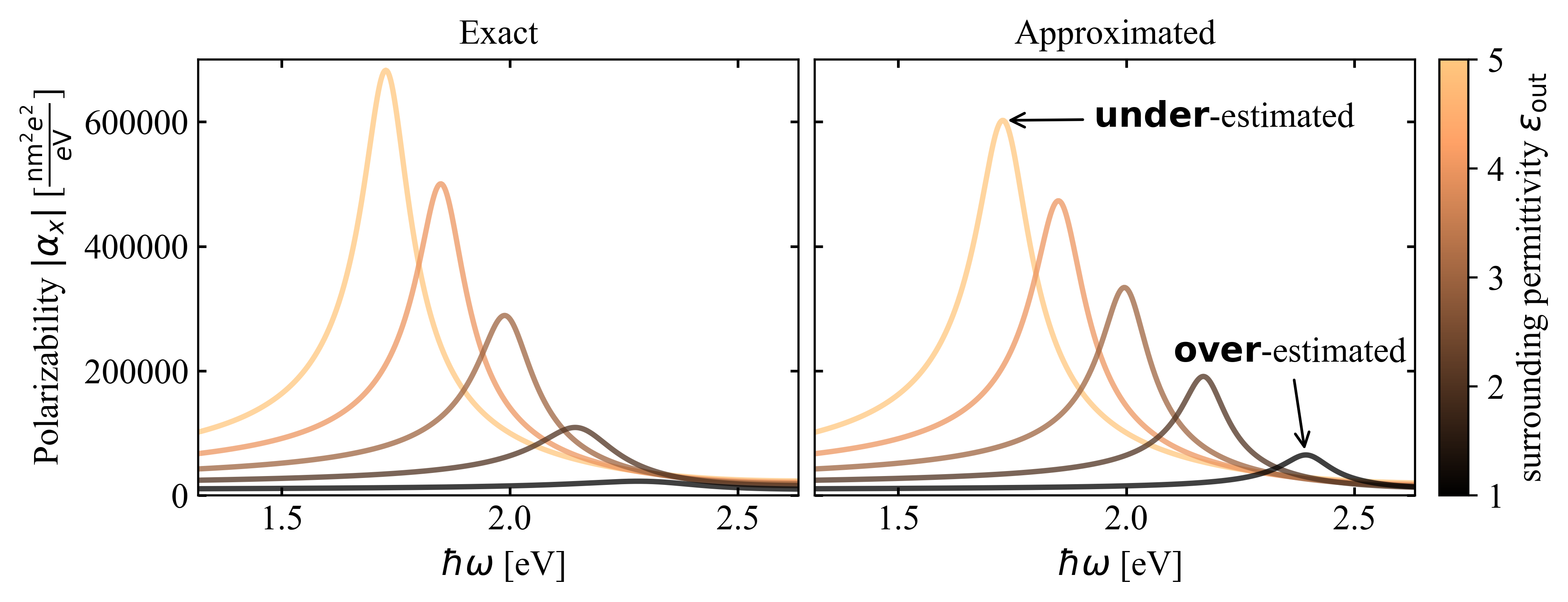}
    \caption{Comparison of exact MNP polarizability $\alpha_x$ (left) and the approximated $\alpha_x$ where $\Im (\varepsilon_b)\approx 0$ (right) for varying surrounding permittivity $\varepsilon_\text{out}$. The $y$-entries of the polarizability are identical; the z-entries here not relevant.}
    \label{app fig: alpha}
\end{figure*}

To test the validity of the approximation $\Im (\varepsilon_b)\approx 0$, we compare the exact polarizability $\mathbcal{\alpha}$ that describes the electric field enhancement around the MNP with the approximated case where the absorption from interband transitions is neglected (i.e., setting $ \Im (\varepsilon_b) = 0$).
Fig.~\ref{app fig: alpha} (left) shows the $x$-entry of the exact polarizability $\alpha_x$, while the right panel depicts the simplified case without interband transition absorption.
In our geometry, the $y$-entry of the polarizability is identical ($\alpha_y = \alpha_x$), and the z-entries not relevant, cp. Appendix \ref{sec: Appendix to 2 dim}.
This way, we assess the approximation for various surrounding permittivities $\varepsilon_\text{out} \in [1,5]$.

The MNP's electric field enhancement, quantified by $\alpha_x$, is a crucial ingredient to the exciton-plasmon coupling strength $g_\text{eff}$.
The comparison reveals that the approximation ($ \Im (\varepsilon_b) = 0$), which is necessary to reduce the description of the MNP dynamics to a single oscillator model, gives rise to an inaccurate dependence on $\varepsilon_\text{out}$ in the simplified coupling, table \ref{table: 3-COM parameters}. Specifically, for small $\varepsilon_\text{out} \approx 1$, the polarizability $\alpha_x$ is significantly overestimated, whereas for large $\varepsilon_\text{out} > 4$, it is underestimated.
Furthermore, the MNP plasmon frequency moves outside the above mentioned allowed frequency range for $\varepsilon_\text{out}\lesssim 2$. 
This leads us to restrict the applicability of the derived coupling strength to $\varepsilon_\text{out} \in [2.5,4.5]$.
Within this range, we observe no significant dependence of the coupling strength $g_\text{eff}$ on the surrounding permittivity $\varepsilon_\text{out}$.
However, in agreement with the theories from Refs.~\cite{greten_strong_2024,salzwedel_spatial_2023}, the minimal field enhancement by the MNP, cp.~Fig.~\ref{app fig: alpha} (left), at $\varepsilon_\text{out} \approx 1$ suggests that higher surrounding permittivities are necessary to reach the strong coupling regime.

\bibliography{bibliography}

\end{document}